\documentclass[lettersize,journal]{IEEEtran}
\usepackage{amsmath,amsfonts}
\usepackage{array}
\usepackage[caption=false,font=normalsize,labelfont=sf,textfont=sf]{subfig}
\usepackage{textcomp}
\usepackage{stfloats}
\usepackage{url}
\usepackage{verbatim}
\usepackage{graphicx}
\usepackage{bm}
\usepackage{multirow}
\usepackage{cite}
\usepackage{mathrsfs}
\usepackage{booktabs}
\usepackage{multirow}
\usepackage{algorithm}
\usepackage{algorithmicx}
\usepackage{algpseudocode}
\usepackage{overpic} 
\usepackage{subfloat} 
\usepackage{bbding}
\usepackage{makecell} 
\usepackage{color,xcolor}
\usepackage{caption}
\usepackage{soul}

\begin{document}

\title{Denoising Designs-inherited Search Framework for Image Denoising}

\author{Zheyu Zhang\thanks{Zheyu Zhang, Yueyi Zhang, and Xiaoyan Sun are with the University of Science and Technology of China, Hefei 230027, China. (e-mail:
 zhangzy0@mail.ustc.edu.cn)}, Yueyi Zhang, Xiaoyan Sun}

\markboth{Journal of \LaTeX\ Class Files,~Vol.~14, No.~8, August~2021}%
{Shell \MakeLowercase{\textit{et al.}}: A Sample Article Using IEEEtran.cls for IEEE Journals}


\maketitle

\begin{abstract}

	How to benefit from plenty of existing denoising designs? Few methods via Neural Architecture Search (NAS) intend to answer this question. However, these NAS-based denoising methods explore limited search space and are hard to extend in terms of search space due to high computational burden. To tackle these limitations, we propose the first search framework to explore mainstream denoising designs. In our framework, the search space consists of the network-level, the cell-level and the kernel-level search space, which aims to inherit as many denoising designs as possible. Coordinating search strategies are proposed to facilitate the extension of various denoising designs. In such a giant search space, it is laborious to search for an optimal architecture. To solve this dilemma, we introduce the first regularization, i.e., denoising prior-based regularization, which reduces the search difficulty. To get an efficient architecture, we introduce the other regularization, i.e., inference time-based regularization, optimizes the search process on model complexity. Based on our framework, our searched architecture achieves state-of-the-art results for image denoising on multiple real-world and synthetic datasets. The parameters of our searched architecture are $1/3$ of Restormer's, and our method surpasses existing NAS-based denoising methods by $1.50$ dB in the real-world dataset. Moreover, we discuss the preferences of $\textbf{200}$ searched architectures, and provide directions for further work.

\end{abstract}


\section{Introduction}
\label{sec:intro}

\IEEEPARstart{I}{mage} denoising is a fundamental and challenging problem in image processing. It aims to restore clean images by removing random noise introduced in image sensing, transmission and processing.  It is not only an important low-level image processing tool, but also a useful pre-processing module for many downstream tasks, such as object detection, image segmentation, and visual analysis.

\begin{figure}[ht]
	\includegraphics[width=\linewidth]{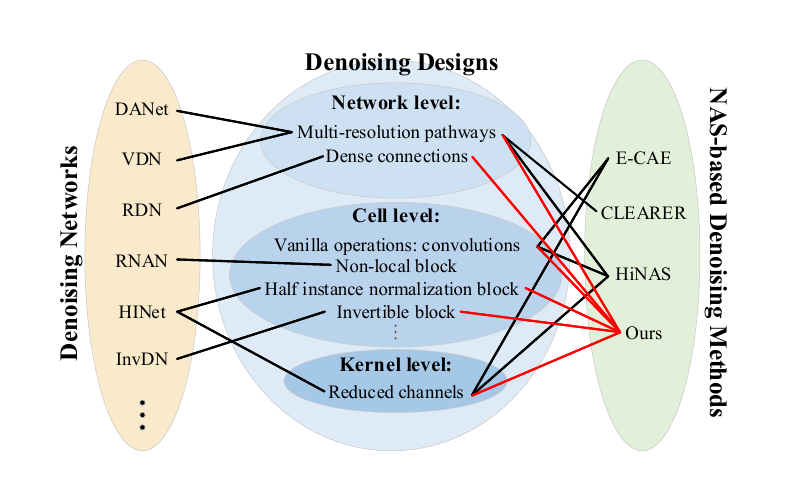}
	\caption{The relationships between denoising designs and denoising networks and NAS-based denoising methods. The denoising designs introduced in the denoising networks are effective based on different perspectives of image denoising. These designs are rarely considered in existing NAS-based denoising methods. We propose the first framework to inherit as many designs as possible.}
	\label{fig:paradigms}
\end{figure}

The field of image denoising is currently dominated by learning-based methods. Deep learning schemes with various denoising operators have been elaborately designed for image denoising. Zhang \textit{et al.}~\cite{zhang2018residual} introduced the non-local attention block in RNAN to capture global information for image denoising. The information-lossless invertible block is presented in InvDN to preserve image details~\cite{liu2021invertible}. 
Chen \textit{et al.}~\cite{chen2021hinet} presented HINBlock to integrate Instance Normalization in the image restoration task.
Multi-Dconv Head Transposed Attention Module~\cite{zamir2022restormer} and Gated block~\cite{zamir2022restormer, chen2022simple} are proposed in transformer-based networks.
Moreover, the dimensions of convolution kernels are adjusted for fast inference \cite{chen2021hinet, wang2022uformer, chen2022simple}.
Network typologies, including multi-resolution pathways~\cite{yue2019variational, zhang2021plug, soh2022variational} and dense connections~\cite{zhang2020residual}\cite{park2019densely}, have also been explored to boost the denoising performance. 
Given all these hand-crafted designs, it is not practical to evaluate the design choices one by one~\cite{chen2022simple}. One cannot help wonder how we can benefit from all these designs effectively for image denoising.

\begin{figure*}[h]
	\centering
	\includegraphics[width=\linewidth]{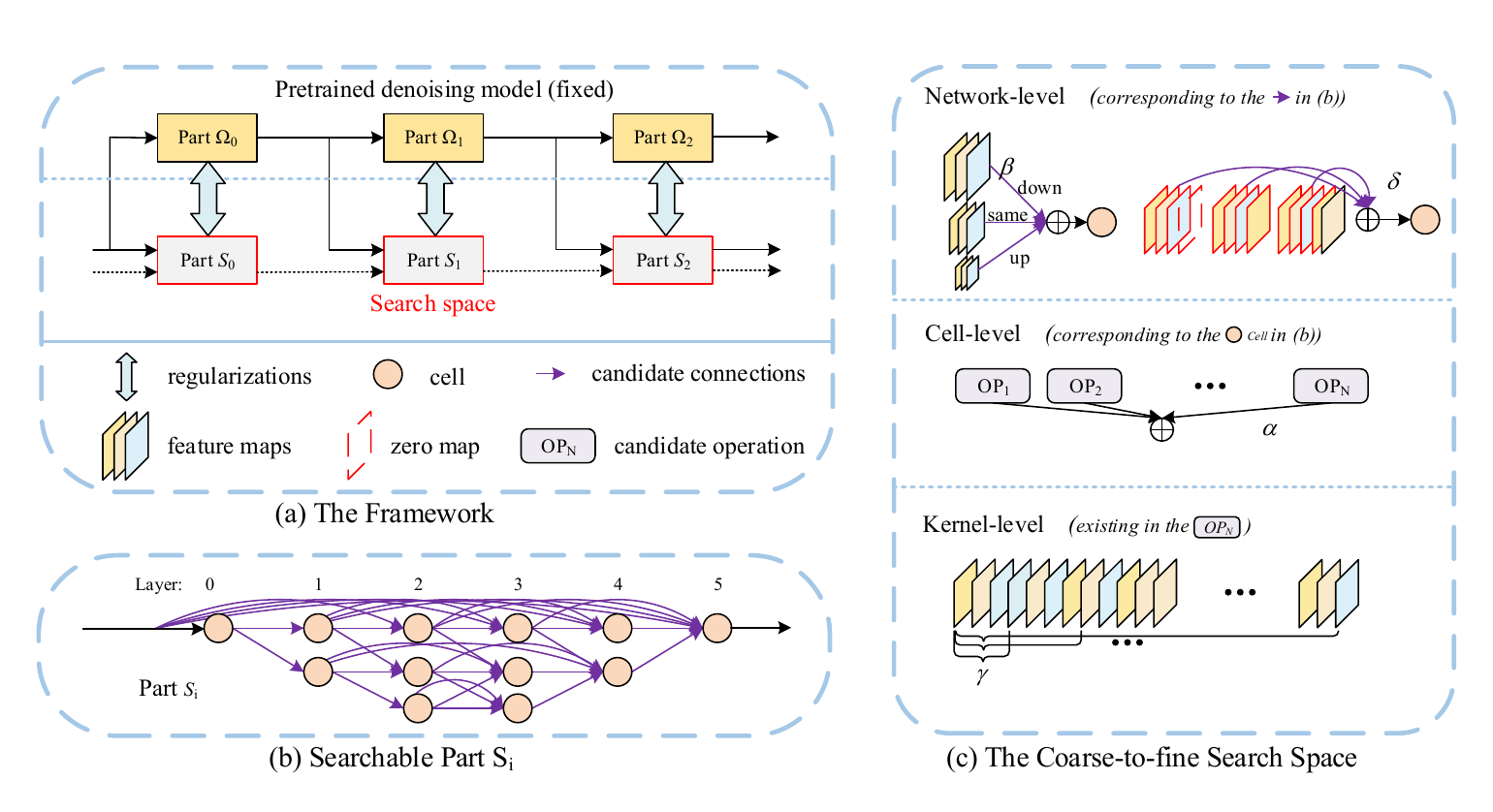}
	\caption{Overview of our proposed method. \textcolor{red}{(a)} Our framework consists of the pretrained denoising model, our search space and two regularizations. Both the model and the space are divided into $3$ individual parts, i.e., $\{\Omega_0(\cdot), \Omega_1(\cdot), \Omega_2(\cdot)\}$ and $\{S_0(\cdot), S_1(\cdot), S_2(\cdot)\}$, respectively. One regularization is the denoising prior-based regularization, which minimizes the distance between $\Omega_{i}(\cdot)$ and $S_i(\cdot)$. The other regularization, which is based on the inference time of each operator, focuses the model complexity. The search process is denoted by the black solid lines. After searching with the regularizations, our searched architecture is obtained by connecting $\{S_0(\cdot), S_1(\cdot), S_2(\cdot)\}$, which is denoted by the black dotted lines. \textcolor{red}{(b)} We enlarge the part $S_i(\cdot)$ that has a coarse-to-fine search space, i.e., the network-level, the cell-level and the kernel-level search space. \textcolor{red}{(c)} We detail the search space that corresponds to part $S_i(\cdot)$. The network-level space corresponding to the arrows in purple modifies the topology. The cell-level space corresponding to the circle selects optimal operator. The kernel-level space corresponding to the operator in the cell prunes redundant channels.}

	\label{fig:framework}
\end{figure*}

Recently, a few efforts have been made to solve this problem via Neural Architecture Search  (NAS) \cite{gou2020clearer, zhang2020memory, zhang2021memory}. Though promising, these schemes explore limited designs, as shown in Fig. \ref{fig:paradigms}, and are hard to extend in terms of search space due to high computational requirements. 
For example, if searching in $8$ operators, i.e., $3\times3$ convolutions with dilation rates of 1/2/3 (conv\_d1/d2/d3), the $3\times3$ convolution with residual connection (conv\_r), skip connection (skip), the invertible block (IB)~\cite{liu2021invertible}, the half instance normalization block (HIN)~\cite{chen2021hinet}, and the swin transformer block (SWIN)~\cite{liang2021swinir}, the search space is giant, which may contain $1e1800$ candidates\footnote{We estimate this based on our search space that is shown in Figure \ref{fig:framework}. The architecture is divided into $3$ parts, each part has $12$ different cells, each cell has $8$ alternative operators, and each operator has $5$ different dimensions. Additionally, the connections between each cell are diverse. For example, for the last cell in one part, the cell has $7$ connected pathways to the other cells, which could generate $( C_7^1+C_7^2+C_7^3+\cdots+C_7^7 ) / 2 \approx 63$ connections. As a result, this cell generates $( 5^8)^{63}\approx 1$e$353$ candidates. In this way, we estimate the candidates are $1$e$1800$.}. 
It would be difficult to search for an optimal architecture in such a space for image denoising.




To remedy these problems, we propose the first search framework that explores the most mainstream designs in image denoising. In our framework, we first propose the coarse-to-fine search space to cover the mentioned designs.
\begin{enumerate}
	\item  \textbf{The network-level space}. The network-level space includes the resolution space and the connection space. The resolution space focuses on the fusion of global but coarse contours and local but fine details. The connection space aims to propagate the necessary details. 
	\item \textbf{The cell-level space}. The cell-level space includes the mentioned mainstream operators. It is flexible and extensible to involve various denoising operators.
	\item \textbf{The kernel-level space}. The kernel-level space explores the dimensions of kernels. It refines the configurations of each convolution.
\end{enumerate}
Then, we propose corresponding search strategies for the coarse-to-fine search space. The network-level search adopts the gradient-based search strategy, which optimizes all pathways at one time. It is compatible with multiple hierarchical features. The cell-level search employs the modified sample-based search strategy that supports the cell-level search space to flexibly extend denoising operators. The kernel-level search also utilizes the sample-based search strategy.

As it is difficult to search an optimal architecture in such a giant search space, i.e., $1e1800$ candidates, we introduce a denoising prior as a regularization for the search space. The denoising prior not only reduces the complexity in searching but also provides experience from the pre-trained denoising model. Meanwhile, to optimize the model complexity, we introduce an inference time regularization for each alternative operator within different configurations, such as the resolution and dimension of input feature maps. In addition, based on the observation of $200$ searched architectures, we draw some conclusions that provide directions for designing a more effective model for image denoising.

The main contributions of this work are summarized as follows:
\begin{itemize}
	\item We introduce the coarse-to-fine search space with compatible search strategies, i.e., the network-level, the cell-level and the kernel-level search, which are the first framework to inherit the merits of mainstream designs in denoising as many as possible.
	\item With the giant search space, denoising prior-based regularization is adopted to reduce the complexity in searching. Inference time-based regularization searches for an efficient model.
	\item Our method outperforms the NAS-based denoising methods and the state-of-the-art methods on multiple synthetic and real-world datasets.
	\item We discuss the preferences of searched architectures, which could provide directions for further work. 
	
\end{itemize}

\section{Related Work}
\label{sec:related}
First, we review the methods of image denoising, including traditional methods, deep learning methods adopting convnet, and methods utilizing the transformer model. Second, we introduce the general NAS methods and NAS-based methods for image denoising.    
\subsection{Image denoising}
Traditional methods utilize hand-crafted priors for image denoising, which are based on the gradient distribution \cite{cho2011image, joshi2009image}, sparsity \cite{dong2011sparsity}, and self-similarity \cite{buades2005non}. These model-driven methods are limited by the assumption of image statistics. Deep learning methods \cite{zhang2017beyond, zhang2018residual, liu2018non, yue2019variational, zamir2020learning, cheng2021nbnet, liu2021invertible, ren2021adaptive, zheng2021deep} benefit from powerful representation and massive data and achieve promising performance. Zhang \textit{et al.} \cite{zhang2017beyond} proposed a one-scale convolution network called DnCNN based on residual learning. Zhang \textit{et al.} \cite{zhang2018residual} explored the effectiveness of non-local attention block in the residual network. Zamir \textit{et al.} \cite{zamir2020learning} proposed a multi-scale network that learns enriched features based on various specific blocks. Cheng \textit{et al.} \cite{cheng2021nbnet} adopted a non-local subspace attention module to project the noisy image into the subspace and to recover the clean image from the subspace. In \cite{liu2021invertible}, Liu \textit{et al.} designed a lightweight denoising network based on the invertible block. In \cite{ren2021adaptive}, Ren \textit{et al.} built a pseudo 3D auto-correlation attention block to extract contextual information from both the channel and spatial dimensions, which is beneficial to image denoising.

Recently, the transformer model, inspired by the Natural Language Processing (NLP), is emerging in Computer Vision (CV). The transformer model is skilled in learning the long-range dependencies that are necessary for image denoising. There are existing excellent works in image restoration, such as IPT \cite{chen2021pre}, SwinIR \cite{liang2021swinir}, Restormer \cite{zamir2022restormer}, Uformer \cite{wang2022uformer} and NAFNet \cite{xiao2022image}. SwinIR~\cite{liang2021swinir} adopts the blocks of the Swin Transformer \cite{Liu_2021_ICCV} and aggregates context information in a shifted window. Restormer and Uformer utilize the designs of U-Net. For each block of the U-Net, Restormer \cite{zamir2022restormer} modifies the self-attention layer for linear complexity and applies the GELU non-linearity activation for complementing details. Uformer \cite{zamir2022restormer} utilizes non-overlapping window-based self-attention to replace the global self-attention and introduces the depth-wise convolutions to better capture the local context. NAFNet \cite{chen2022simple} replaces the GELU with the proposed SimpleGate and channel attention, and simplifies the blocks of Restormer. In summary, various denoising blocks, which are based on different priors, are devised in the above mentioned denoising methods. However, less attention is given to carrying multiple specific blocks in one architecture.

\subsection{Neural architecture search}
\label{sec:related_NAS}
Neural architecture search is proposed for automatically designing architectures for the target task. It can be categorized into $3$ dimensions \cite{elsken2019neural}: search space, search strategy, and performance estimation strategy. The search space defines all potential candidates. The search strategy represents how to explore the search space. The performance estimation strategy evaluates the searched architecture.  

For the search space, in earlier works \cite{real2017large, zoph2016neural}, researchers adopted the chain structure that is written as a fixed sequence with multiple layers. The methods search for suitable operators in each layer. Then, a few works \cite{liu2018darts,dong2019searching} focused on searching for repeatable cells instead of different layers. The outer predefined network structure consists of repeatable cells that contain multiple possible layers. More recently, because the predefined network structure does not match the application for different tasks, researchers have tried more flexible search spaces to promote the performance of their searched architectures \cite{liu2019auto, chen2019fasterseg, fang2020densely, zhang2021dcnas}. However, their architectures neglect the importance of multiple resolutions. Specifically, the methods \cite{liu2019auto, chen2019fasterseg} decode the architecture with only one pathway from multi-resolution pathways. Besides, the modules \cite{liu2019auto, chen2019fasterseg, zhang2021dcnas} for down-sample/up-sample neglect the image texture, which is necessary for denoising.

More researchers have focused on the search strategy, which has a significant influence on performance and efficiency. Frequent search strategies are built on various algorithms, including evolutionary algorithms \cite{real2017large, suganuma2018exploiting, song2020efficient, li2020all}, reinforcement learning \cite{zoph2016neural, pham2018efficient} and gradient-based methods\cite{liu2018darts, liu2019auto, chen2019fasterseg, xie2018snas, hu2020dsnas, dong2019searching}. Evolutionary algorithm-based methods first initialize a set of models and then evolve to obtain better architectures. Reinforcement learning-based methods focus on how to represent and optimize the agent's policy. Zoph \textit{et al.} \cite{zoph2016neural} utilized the accuracy of the `child network' to compute the policy gradient, which updates the RNN-controller. Thus, the RNN-controller has higher possibilities of generating the new `child network', which receives high accuracy. Baker \textit{et al.} \cite{baker2016designing} took Q-Learning to train the agent and sequentially chose CNN layers. Li \textit{et al.} \cite{liu2018darts} proposed DARTS, a gradient-based method, which relaxes the discrete search space to be continuous. Specifically, DARTS builds a directed acyclic graph composed of nodes and edges. Nodes are connected by edges that represent multiple operators. The operators are assigned with continuous architecture weights. The representation of a node is the weighted summation of the outputs of these operators. With gradient descent, the final discrete architecture is derived based on the maximum architecture weights. Compared with NAS methods based on evolutionary algorithms or reinforcement learning, gradient-based methods are much faster and more economical. However, the methods based on DARTS are unfriendly to GPU memory. As DARTS has to save tensors and gradients of all operators, its GPU memory usage proliferates with an increasing search space. 

E-CAE~\cite{suganuma2018exploiting} is the first NAS-based denoising method that adopts evolutionary search. It searches an architecture based on vanilla convolutions with different kernels. This method is more time-consuming than the gradient-based denoising methods \cite{zhang2020memory, zhang2021memory, gou2020clearer, ning2021searching}. These methods employ DARTS as their search strategies. HiNAS \cite{zhang2020memory, zhang2021memory} searches repeatable cells and utilizes a layer-wise architecture-sharing strategy to improve the diversity of cells. CLEARER \cite{gou2020clearer} builds a chain-structure space to search cells that consist of predefined modules. However, as shown in Figure \ref{fig:paradigms}, these methods only explore the limited search space and neglect the importance of specific denoising operators. In addition, NAS-based denoising methods should be flexible to extend the new operators into the search space. As an increasing number of operators based on different perspectives are proposed, it is necessary to extend the operators into the search space without increasing the burden of GPU memory. This encourages us to propose a feasible framework to flexibly extend specific denoising operators into the search space.


\section{Method}
\label{sec:method}
In this section, we first provide an overview of our framework. Then, we present the network-level, the cell-level, and the kernel-level search. After that, two regularizations are illustrated. Finally, we describe our search pipeline.

\subsection{The Framework}
\label{sec:overview}
As shown in Figure \ref{fig:framework}a, our framework contains the search space, the pretrained denoising model and two regularizations. Our search space provides diverse alternative architectures that inherit denoising designs. The pretrained denoising model serves as the supervision for searching, which is designed for the following regularization. Two regularizations are utilized to trade off the model performance against the model complexity. The prior-based regularization minimizes the distance of features between the pretrained denoising model and the architectures in our search space. Specifically, both the pretrained denoising model and our search space are divided into several individual parts, i.e., $\Omega_i(\cdot)/S_i(\cdot)$. The pretrained part $\Omega_i(\cdot)$ serves as the prior to supervise the performance of the searchable part $S_i(\cdot)$. More importantly, denoising prior-based regularization reduces the search complexity. Assume that the search complexity for each part is $\mathcal{O}(m)$, where $m$ is the total number of candidates in the search process for each individual part $S_i(\cdot)$. The denoising prior-based regularization reduces the search complexity from $\mathcal{O}(m^3)$ to $\mathcal{O}(m+m+m)$, i.e., $\mathcal{O}(m)$. The other regularization, i.e., inference time-based regularization, optimizes the model complexity. It builds a lookup table for measuring the average inference time of different settings of denoising operators. The model efficiency can be measured by the weighted summation of the inference time of the selected operators.  With the two regularizations, our searchable parts have the capability for parallelly searching the optimal architecture for image denoising. The final searched architecture is implemented by connecting our searched parts together. In our experiment, we choose $3$ individual parts for the search space.

\subsection{The Coarse-to-fine Search}
\label{sec:Coarse-to-fine}
As shown in Figure \ref{fig:framework}b, each part has a complete coarse-to-fine search space, which guarantees the diversity of our alternative architectures. Our search space contains the network-level, the cell-level, and the kernel-level space, as shown in Figure \ref{fig:framework}c. The network-level search considers the external connections of multiple cells. The cell-level search aims to find the optimal operator in each cell. In each cell, the kernel-level search provides balanced widths for the selected operators. 

\textbf{The network-level search:} The network-level search seeks the optimal network topology for image denoising. 
Topologies with multi-resolutions and dense connections have proven their effectiveness in denoising methods~\cite{yue2019variational, zhang2021plug, soh2022variational, zhang2020residual}. In our work, we extend these two topologies to our search space.
The multi-resolution topology is described as a network cell that takes multi-resolution feature maps and generates an aggregated feature map. However, processing multi-resolution feature maps for each cell is computationally demanding and unnecessary. The feature maps at various resolutions should be selectively sent to the network cell. To this end, we propose our searchable multi-resolution process, replacing the fixed multi-resolution process.

\begin{table*}[tp]
	\centering
	\caption{The effects of different down-sample/up-sample modules for image denoising. The models are evaluated on the `Case 1 Set5' dataset. `Concatenation': We assume the feature map are down-sampled at the ($m+1$)-th layer, i.e., the third layer in the toy model. The final up-sampled feature map is concatenated with the $m$-th feature map. The concatenated feature maps are sent into a convolution for decreasing the channel dimension.}
	\begin{tabular}{@{}lccccccc@{}}
	\toprule
	\multirow{4}{*}{Models} & \multicolumn{2}{c}{Down-sample} & \multicolumn{4}{c}{Up-sample}                        & \multirow{4}{*}{PSNR}  \\ \cmidrule(r){2-3}\cmidrule(r){4-7}
	 & Averagepooling  & Convolution-stride2  & Bilinear & Transposed convolution & Pixelshuffle & Concatenation &       \\
	 &				\begin{minipage}[b]{0.15\columnwidth}\raisebox{-.5\height}{\includegraphics[width=\linewidth,trim=20 25 10 25,clip]{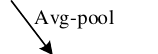}}\end{minipage}              
	 				& \begin{minipage}[b]{0.15\columnwidth}\raisebox{-.5\height}{\includegraphics[width=\linewidth,trim=20 25 10 25,clip]{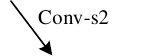}}\end{minipage}         
					& \begin{minipage}[b]{0.15\columnwidth}\raisebox{-.5\height}{\includegraphics[width=\linewidth,trim=20 25 10 25,clip]{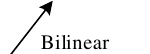}}\end{minipage}           
					& \begin{minipage}[b]{0.15\columnwidth}\raisebox{-.5\height}{\includegraphics[width=\linewidth,trim=20 25 10 25,clip]{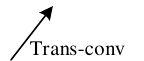}}\end{minipage}    
					&  \begin{minipage}[b]{0.15\columnwidth}\raisebox{-.5\height}{\includegraphics[width=\linewidth,trim=20 25 10 25,clip]{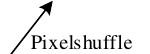}}\end{minipage}         
					&  \begin{minipage}[b]{0.15\columnwidth}\raisebox{-.5\height}{\includegraphics[width=\linewidth,trim=10 25 10 25,clip]{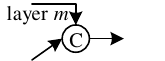}}\end{minipage}             &       \\ \midrule
	Model-1 & \Checkmark			&               &  \Checkmark        &            			&             			 &               & 24.83 \\
	Model-2 & \Checkmark			&               &  \Checkmark        &            			&           		     &  \Checkmark   & 28.66 \\
	Model-3 & &  \Checkmark   &  \Checkmark        &            			&             			 &  \Checkmark   & 28.96 \\
	Model-4 & &  \Checkmark   &  			         &  \Checkmark          &              			 &  \Checkmark   & 28.35 \\
	Model-5 & &  \Checkmark   &  			         &            			&    \Checkmark          &  \Checkmark   & \textbf{29.12} \\ 
	\bottomrule
	\end{tabular}
	\label{tab:up-down}
	\end{table*}

Before introducing the searchable multi-resolution process, we first investigate the up-sample and down-sample modules, which are the essential elements for the multi-resolution process. The different up-sample/down-sample modules have distinct effects on image denoising. We demonstrate a toy experiment to compare the performance of various up-sample/down-sample modules for image denoising, and select the optimal pair. The experiment is based on a model that is sequentially constructed by $2$ vanilla convolutions, a down-sample module, $12$ vanilla convolutions, an up-sample module, and $2$ vanilla convolutions. A long skip connection for residual learning adds the input to the final convolution output. For down-sampling, we select the convolution with stride 2 and the average pooling. For up-sampling, the bilinear, transposed convolution, and the pixelshuffle are alternative modules for comparison. 

As Table \ref{tab:up-down} shows, from the comparison between `Model-2' and `Model-3', the down-sampling by convolution with the stride 2 is better than average pooling. This is mainly because the convolution provides adaptive features. Comparing `Model-2' with `Model-1', we can see that the up-sample module heavily impacts the performance of image denoising, especially when it is combined with concatenation. This improves the performance by 4 dB. In addition,  `Model-5' achieves the best performance. Hence, we adopt the convolution with the stride 2 as our down-sample module and the pixelshuffle with the concatenation as our up-sample module. 

After determining the up-sampling and down-sampling modules, we design the search strategy for the multi-resolution search. We utilize the architecture weight $\bm{\beta}$ to represent the probability of selected resolution and determine the pathways of connected cells. Based on $\bm{\beta}$ and the down-sample/up-sample modules, the aggregation of multi-resolution feature maps can be described as follows:
\begin{align}
	{\boldsymbol O}_{RES}&={\overline\beta}_{\frac s2\rightarrow s,\;l+1}{\text{f}}_{up}({\boldsymbol x}_{\frac s2,\;l},\;{\boldsymbol x}_{s,\;m})+{\overline\beta}_{s\rightarrow s,\;l+1}{\boldsymbol x}_{s,\;l} \notag \\
	&+{\overline\beta}_{2s\rightarrow s,\;l+1}{\text{f}}_{down}({\boldsymbol x}_{2s,\;l}),
	\label{eq:resolution}
\end{align}
where $O_{RES}$ is the aggregated multi-resolution features at the ($l+1$)-th layer, and ${\overline\beta}_{\frac s2\rightarrow s}$ is the normalized scalar, which is implemented by $\bm{{\overline\beta}}=\text{softmax}(\bm{\beta})$. Notably, the function $\text{f}_{up}(\cdot)$ is implemented by our up-sample module, i.e., the combination of the pixelshuffle and the concatenation. 
The feature map ${\boldsymbol x}_{\frac s2,\;l}$ is the output of the $l$-th layer at the $\frac s2$ resolution. The feature map ${\boldsymbol x}_{s,\;m}$ is the output of the $m$-th layer at the $s$ resolution. The $m$-th layer is the last layer before down-sampling. The pixelshuffle takes the feature map ${\boldsymbol x}_{\frac s2,\;l}$ and its output is concatenated with the feature map ${\boldsymbol x}_{s,\;m}$.


After training $\bm{\beta}$, we utilize Breadth First Search algorithm (BFS) \cite{chen2019fasterseg, zhang2021dcnas} to decode the topology of the architecture from the last cell to the first cell. Meanwhile, we set a threshold to select multiple resolutions, because some cells may need more than one-resolution feature map. If the maximum ${\overline\beta_i}>0.5,\; \text{where}\; {\overline\beta_i} \in \{{\overline\beta}_{\frac s2\rightarrow s}$, ${\overline\beta}_{s\rightarrow s}, {\overline\beta}_{2s\rightarrow s}\}$, we select one resolution path. Otherwise, we select more resolution paths until $\sum_i {\overline\beta}_i>0.5$. In this way, the cell could select features from more than one resolution for aggregating the global contours and the local textures.


Except for the resolution search, we introduce the dense connection search to decrease the loss of information in image denoising, which is inspired by RDN~\cite{zhang2020residual}, the Residual Dense Network. One contribution of RDN is that it introduces dense connections. However, too many skip connections affect the efficiency of inference. Consequently, we propose selective dense connection to preserve the essential low-level features. We utilize the architecture weight $\bm{\delta}$ to represent the importance of corresponding connections. The aggregation of dense connected feature maps is implemented as follows::
\begin{equation}
	{\boldsymbol O}_{CONN}=\sum_{l=0}^{\Vert \bm{\delta} \Vert_0-1} {\overline{\delta}}_{l}{\boldsymbol x}_{l},
	\label{eq:connect}
\end{equation}
where $\bm{\overline\delta}=\mathrm{softmax}(\bm{\delta})$. $l \in [0, {\Vert \bm{\delta} \Vert_0}-1]$, and ${\Vert \bm{\delta} \Vert_0}$ is the total number of cells in the current resolution. We preserve the multiple skip connections based on $\bm{\delta}$, i.e., $\sum_i {\overline\delta}_i>0.5$. It should be noted that the channels of multiple feature maps are different, resulting in these feature maps not being summed. This problem is overcome by coordinating with our kernel-level search. Given the characteristic of our kernel-level search, the feature maps of the former channels are more significant. Based on this, we first assume the unified channels that are determined by the output dimensions of the last cell. If the channels of connected feature maps are less than the unified channels, we pad zero map to the connected feature maps until their channels are equal to the unified channels. Otherwise, we select the former channels of connected feature maps. In this way, the feature maps that come from different skip connections can be summed. Thus, we select the most important multiple skip connections for preserving the necessary information. 

\begin{table*}[htp]
	\centering
	\caption{The difference of various search strategies for the cell-level search.}
	\begin{tabular}{@{}cccc@{}}
	\toprule
					  & Forward (The cell output) & Backward (The cell gradient) & References        \\ \hline
	\multirow{2}{*}{Eqs. (\ref{eq:darts}) (\ref{eq:gra_darts})} & ${\boldsymbol O}_{Cell} = \sum_{i=0}^{{\Vert \bm{\alpha} \Vert_0}-1} \overline{\alpha}_i O_i(x)$         & $\frac{\partial \mathcal{L}}{\partial \alpha_i} = \frac{\partial \mathcal{L}}{\partial y} \overline{\alpha}_i \left[\sum_{m=0, m\neq i}^{{\Vert \bm{\alpha} \Vert_0}-1}(O_i(x)-O_m(x))\overline{\alpha}_m\right] $         & \multirow{2}{*}{\begin{tabular}[c]{@{}c@{}}\cite{liu2018darts}, adopted in other\\ denoising methods \cite{zhang2020memory, zhang2021memory, gou2020clearer}.\end{tabular}} \\ \cline{2-3}
					  & calculate \textbf{all} alternative operators.         & calculate \textbf{all} alternative operators.         &                   \\ \hline
	\multirow{2}{*}{Eqs. (\ref{eq:gdas1}) (\ref{eq:gra_gdas})} & ${\boldsymbol O}_{Cell} = \sum_{i=0}^{{\Vert \bm{\alpha} \Vert_0}-1}z_i O_i(x)$        & $\frac{\partial \mathcal{L}}{\partial \alpha_i} = \frac{\partial \mathcal{L}}{\partial y}\frac{{\widetilde z}_i}{\lambda_{tem}\alpha_i}\left[O_i(x)-\sum_{m=0, m\neq i}^{{\Vert \bm{\alpha} \Vert_0}-1}O_m(x){\widetilde z}_m\right]$          & \multirow{2}{*}{\cite{dong2019searching, jang2016categorical, xie2018snas}} \\ \cline{2-3}
					  & calculate \textbf{one} operator.        & calculate \textbf{all} alternative operators.         &                   \\ \hline
	\multirow{2}{*}{Eqs. (\ref{eq:gdas1}) (\ref{eq:gra_final})} & ${\boldsymbol O}_{Cell} = \sum_{i=0}^{{\Vert \bm{\alpha} \Vert_0}-1}z_i O_i(x)$        & $\frac{\partial \mathcal{L}}{\partial \alpha_i}=\frac{\partial \mathcal{L}}{\partial y} O_i(x) \nabla_{\alpha_i}{\rm log} \, \frac{{\rm exp}(\alpha_i)}{\sum_{j=0}^{{\Vert \bm{\alpha} \Vert_0}-1}{\rm exp}(\alpha_j)}$          & \multirow{2}{*}{\cite{maddison2016concrete, hu2020dsnas}} \\ \cline{2-3}
					  & calculate \textbf{one} operator.        & calculate \textbf{one} operator.         &                   \\ 
	\bottomrule
	\end{tabular}
	\label{tab:strategies}
\end{table*}

\textbf{The cell-level search:} After the effective topology is determined by the network-level search strategy, we need to fill the topology with effective cells, which is illustrated in Figure \ref{fig:framework}b. The common cell-level search space \cite{zhang2020memory, gou2020clearer, ning2021searching} consists of multiple convolutions with different kernel sizes and dilation rates. We add specific denoising operators, which have been evaluated in existing hand-crafted networks \cite{liu2021invertible, liang2021swinir, chen2021hinet}, into our cell-level search space. Specifically, they are the invertible block \cite{xiao2020invertible, liu2021invertible, huang2022winnet}, the half instance normalization block \cite{chen2021hinet}, and the swin transformer block \cite{Liu_2021_ICCV, liang2021swinir}.
The invertible block emphasizes the lossless information, which preserves the beneficial features for denoising. The half instance normalization block utilizes statistical information to enhance the features. The swin transformer block models long-range dependency with the shifted window scheme, which is different from the alternative operator of convolutions. 

\begin{figure}[tp]
	\centering
	\includegraphics[width=0.9\linewidth]{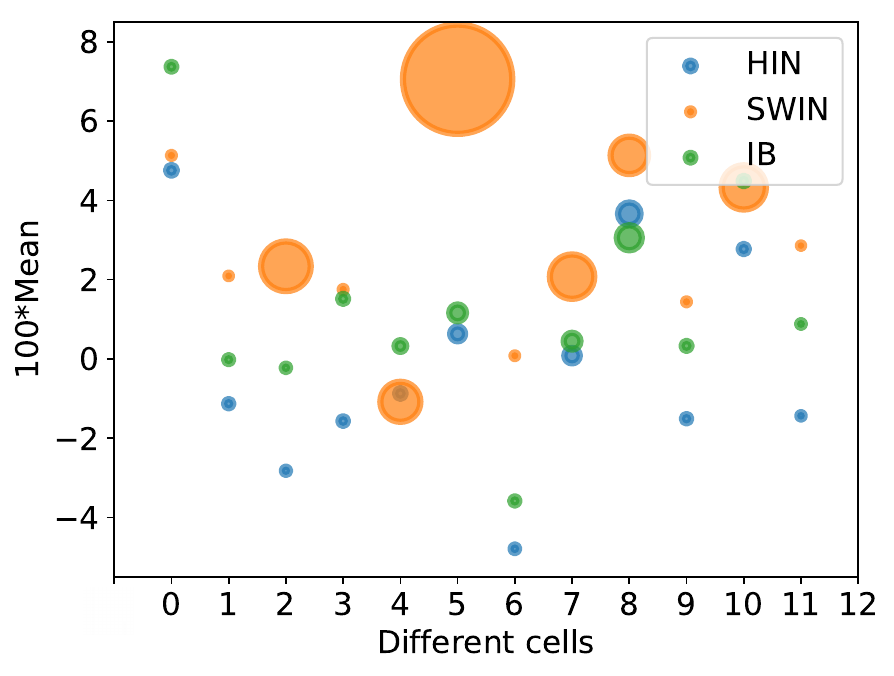}
	\caption{The feature statistics of specific denoising operators, i.e., `HIN', `SWIN', and `IB'. The position of each circle represents the index and the mean of a cell. The radius of each circle represents the standard deviation.}
	

	\label{fig:feat_mean_std}
\end{figure}

Our cell-level search space consists of the $3\times3$ convolutions with dilation rates of 1/2/3 (conv\_d1/d2/d3), the $3\times3$ convolution with residual connection (conv\_r), skip connection (skip), the invertible block (IB), the half instance normalization block (HIN), and the swin transformer block (SWIN). Notably, the alternative operator, i.e., `skip', adjusts the depth of our selected architecture.
Specifically, it generates the features without any process. In Figure~\ref{fig:framework}b, since our architecture predefines the depth, i.e., the layer number, the `skip' could serve as a placeholder in the predefined cell. If the layer selects the `skip' operator, it can be viewed as a decrease in the depth of the selected architecture. In this way, the depth of our architecture is adjustable. 

Based on the cell-level search space, we first utilize the statistics of features to verify that the denoising operators emphasize different perspectives of image denoising. We train our `Part $S_i$' until convergence. In each cell, we calculate the statistics of features generated by the denoising operators. As these operators are in the same place, the statistics should be distinct if the denoising operators emphasize different perspectives. As shown in Figure~\ref{fig:feat_mean_std}, `SWIN' has a wider radius which is caused by the long-range dependency, while `IB' and `HIN' have a limited receptive field, which leads to a narrow radius.

With the 8 alternative operators in the cell-level search space, it is infeasible to adopt the gradient-based search strategy\cite{liu2018darts, zhang2020memory, ning2021searching} like Eqs. (\ref{eq:resolution}) (\ref{eq:connect}) due to the giant GPU memory usage. We analyze the reasons by the forward and backward propagation.

In the gradient-based search strategy, the cell can be calculated as: 
\begin{equation}
	\label{eq:darts} 
	{\boldsymbol O}_{Cell} = \sum_{i=0}^{{\Vert \bm{\alpha} \Vert_0}-1}\frac{{\rm exp}(\alpha_i)}{\sum_{j=0}^{{\Vert \bm{\alpha} \Vert_0}-1} {\rm exp}(\alpha_j)} O_i({\boldsymbol x}) 
	= \sum_{i=0}^{{\Vert \bm{\alpha} \Vert_0}-1} \overline{\alpha}_i O_i({\boldsymbol x}),
\end{equation}   
where ${\Vert \bm{\alpha} \Vert_0}$ is the total number of our alternative operators and $O_i(\cdot)$ is the operator. The $\alpha_i$ is the architecture weight of each operator. The $\bm{\overline{\alpha}}=\text{softmax}(\bm{\alpha})$ represents the normalized tensor.
In the backward propagation, with the surrogate loss $\mathcal{L}$, the gradient for the weight $\alpha_i$ is: 
\begin{align}
	\label{eq:gra_darts}
	\frac{\partial \mathcal{L}}{\partial \alpha_i} 
	= \frac{\partial \mathcal{L}}{\partial y} \overline{\alpha}_i \left[\sum_{m=0, m\neq i}^{{\Vert \bm{\alpha} \Vert_0}-1}(O_i({\boldsymbol x})-O_m({\boldsymbol x}))\overline{\alpha}_m\right]. 
\end{align}

From Eqs. (\ref{eq:darts}) (\ref{eq:gra_darts}), we find that the tensors flow through all operators in both the forward and backward propagations, which consumes too much GPU memory. The authors in \cite{dong2019searching} adopt the `arg max' function in the forward propagation to avoid traversing all the possibilities of operators. Here we utilize it to relax the GPU memory. Meanwhile, the Gumbel softmax trick \cite{jang2016categorical} is introduced to relax the `arg max' function.
The process for the forward propagation can be expressed as:

\begin{equation}
	\label{eq:gdas1}
	{\boldsymbol O}_{Cell} = \sum_{i=0}^{{\Vert \bm{\alpha} \Vert_0}-1}z_i O_i({\boldsymbol x}),
\end{equation}
where the vector $[z_0, \cdots ,z_i, \cdots, z_{{\Vert \bm{\alpha} \Vert_0}-1}]$ is a one-hot vector, which can be derived from $\rm{one\_hot}(\mathop{\arg\max}\limits_{i}(\alpha_i+ G_i))$. The $G_i$ is i.i.d samples drawn from Gumbel (0,1). In the forward propagation, the one-hot vector is implemented by the `arg max' function. However, the `arg max' function is non-differentiable. Based on the Gumbel softmax trick, in the backward propagation, we adopt the $\widetilde{z_i}$ to reparameterize $z_i$. The $\widetilde{z_i}$ is written as:
\begin{equation}
	\label{eq:gum}
	\widetilde{z_i} = \frac{{\rm exp}((log(\alpha_i)+G_i)/\lambda_{tem})}{\sum_{j=0}^{{\Vert \bm{\alpha} \Vert_0}-1} {\rm exp}((log(\alpha_j)+G_j)/\lambda_{tem})},
\end{equation}
where $\lambda_{tem}$ is the temperature. The gradient is derived as follows: 
\begin{align}
	\label{eq:gra_gdas}
	\frac{\partial \mathcal{L}}{\partial \alpha_i} 
	=\frac{\partial \mathcal{L}}{\partial y}\frac{{\widetilde z}_i}{\lambda_{tem}\alpha_i}\left[O_i({\boldsymbol x})-\sum_{m=0, m\neq i}^{{\Vert \bm{\alpha} \Vert_0}-1}O_m({\boldsymbol x}){\widetilde z}_m\right],
\end{align}
Since the ${\widetilde z}_m \neq 0$, the features generated by $O_m(\cdot)$ have to be preserved, indicating that we have to save not only the selected operator $O_i(\cdot)$  in the GPU memory for the back-propagation, but also the other operators $O_m(\cdot)$.

As we want to extend our cell-level search space from existing denoising operators as many as possible, it is non-trivial to reduce the GPU memory in the backward propagation. A better way is to back-propagate gradients from the sampled operator. Therefore, we finally adopt the strategy in \cite{maddison2016concrete, hu2020dsnas}. The forward propagation is the same as Eq. (\ref{eq:gdas1}), and the backward propagation is modified as:
\begin{equation}
	\label{eq:gra_final}
	\frac{\partial \mathcal{L}}{\partial \alpha_i} = \frac{\partial \mathcal{L}}{\partial y} O_i(x) \nabla_{\alpha_i}{\rm log} \, \frac{{\rm exp}(\alpha_i)}{\sum_{j=0}^{{\Vert \bm{\alpha} \Vert_0}-1}{\rm exp}(\alpha_j)},
\end{equation}
the $O_m(\cdot)$ is divided out and only the selected operator $O_i(\cdot)$ is preserved. The GPU memory is further saved by discarding the gradient that is calculated by the other operators. Therefore, this search strategy can easily extend additional denoising operators without worrying about the GPU memory usage. To better understand the differences among these three search strategies, we draw a table to illustrate, as shown in Table \ref{tab:strategies}.

\textbf{The kernel-level search:} After determining the operator for each cell, we investigate the optimal width for convolutions in the operator for a more efficient architecture for image denoising. 

Based on the slimmable convolution \cite{yu2019slimmable, chen2019fasterseg}, we search the kernel width of each cell. First, we predefine a vector ${\boldsymbol R}=[\frac{8}{8}, \frac{7}{8}, \frac{6}{8}, \frac{5}{8}, \frac{4}{8}]$ and the weight of a convolution kernel written as $W[0:C_{in}, 0:C_{out}]$. The slimmable convolution indicates that we can obtain the sub-kernel with the slimmable upper bound ${\boldsymbol R}[i] \times C$, i.e., $W_{sub}=W[0:\boldsymbol R[i]\times C_{in}, 0:\boldsymbol R[j]\times C_{out}]$. Then, similar to the search strategy in Eqs. (\ref{eq:gdas1}) (\ref{eq:gra_gdas}), the searched kernel can be expressed as:
\begin{equation}
	W^* = W[0:\boldsymbol R[i_{l-1}]\times C_{in}, 0:\boldsymbol R[i_l]\times C_{out}],
\end{equation}
where $i_l=\mathop{\arg\max}\limits_{i}(\gamma_{i,l}+ G_{i,l})$, and the $\bm{\gamma}$ is the architecture weight for the kernel-level search. The features generated by the searched kernel is:
\begin{equation}
	{\boldsymbol O}_{KRN}=W^* * {\boldsymbol x}.
	\label{eq:kernel}
\end{equation}

In this way, the minimum sub-kernel extracts the most fundamental features, and the kernel in the subsequent channels enhances the fundamental features. This encourages us to sum the feature maps in the fundamental channels. Coordinated with the strategy in the kernel-level search, we solve the problem of incompatible channels in the dense connection search.

\textbf{Motivations of search strategies:} We adopt different search strategies for the coarse-to-fine search space in light of the characteristics. In the network-level search, we want to preserve the multiple pathways of the searched architecture. A more economical way is to train all pathways together at one time. Therefore, we adopt the DARTS \cite{liu2018darts} -like strategy to train the multiple pathways simultaneously. In the cell-level search, we aim to cover as many specific denoising operators as possible, indicating that our cell-level search space is huge. 
For updating architecture weight for each operator, it is impossible to save and calculate tensors and gradients of all operators.
As a result, we adopt the search strategy of Eqs. (\ref{eq:gdas1}) (\ref{eq:gra_final}) to guarantee that only the sampled operator is trained. In the kernel-level search, as the features generated by sub-kernels have unequal channels, they cannot be summed. We avoid the summation of the DARTS-like strategy by the sampling strategy. 
Although both sampling strategies of Eqs. (\ref{eq:gdas1}) (\ref{eq:gra_gdas}) and Eqs. (\ref{eq:gdas1}) (\ref{eq:gra_final}) are feasible, the latter sampling strategy pays more attention on the reduction of GPU memory usage and this is not necessary for limited space for the kernel-level search. We adopt the strategy Eq. (\ref{eq:kernel}), which is similar to the sampling strategy Eqs. (\ref{eq:gdas1}) (\ref{eq:gra_gdas}). 

\subsection{Regularization of The Search Space}
\label{sec:regularization}
\textbf{Denoising Prior-based Regularization:} 
We propose denoising prior-based regularization to reduce the search difficulty due to the giant search space. 
As shown in Figure \ref{fig:framework}a, we divide the search space into $3$ parts, i.e., $\{S_0(\cdot), S_1(\cdot), S_2(\cdot)\}$. To ensure that each part can separately search an optimal architecture, it is non-trivial to provide a supervision for each part. We adopt a pretrained denoising model as a prior-based regularization, which can also be divided into $3$ pretrained parts, i.e., $\{\Omega_0(\cdot), \Omega_1(\cdot), \Omega_2(\cdot)\}$. We minimize the distance of features between $S_i(\cdot)$ and $\Omega_i(\cdot)$. This can be formulated as follows:
\begin{equation}
	\mathcal{L}_{dp} = ||S_i({\boldsymbol x})-\Omega_i({\boldsymbol x})||_2^2,
	\label{eq:dp}
\end{equation} 
where ${\boldsymbol x}$ is the input for each part. For the part $S_0({\boldsymbol x})$, ${\boldsymbol x}$ is the noisy image. For the parts $\{S_1({\boldsymbol x}), S_2({\boldsymbol x})\}$, $x$ is the features delivered from the pretrained parts $\{\Omega_0({\boldsymbol x}), \Omega_1({\boldsymbol x})\}$. This ensures that the part $S_i(\cdot)$ in the search space can be searched separately. After all parts are searched, the final complete architecture is implemented by connecting these parts together. 

This regularization reduces the complexity in searching by inheriting the experience of excellent pretrained model. Besides, it provides feasibility to search an architecture in parallel.

\textbf{Inference time-based Regularization:} Except the denoising prior-based regularization that supervises the search process on model performance, we propose the inference time-based regularization to supervise the search process on model complexity. 
Searching for an efficient architecture is necessary, especially for edge-platform deployments. 
However, it is rarely considered in existing NAS-based denoising methods. The authors in CLEARER \cite{gou2020clearer} introduced a complexity regularization. However, the regularization only counts the number of two proposed modules. These two types of modules have different complexities. As the total number of modules is fixed, the architecture will be more efficient with more light module. This regularization cannot be used with many operations.
We introduce the complexity regularization based on the inference-time lookup table.

We count the inference time of each operator with different settings, i.e., the operator $\alpha_i$ with different resolution $\bm{\beta}$ and width $\bm{\gamma}$ of features. We count them in $1000$ times and generate an average inference time $t(\alpha_i|\bm{\beta}, \bm{\gamma})$ which is saved on the lookup table. $t(\alpha_i|\bm{\beta, \gamma})$ indicates $\alpha_i$ is the variable, but $\bm{\beta, \gamma}$ are fixed and uniformly initialized. The complexity regularization for $\bm{\alpha}$ can be formulated as:
\begin{equation}
	\mathcal{L}_{\bm{\alpha}} = \sum_{i=0}^{\Vert \bm{\alpha} \Vert_0-1}\alpha_i t(\alpha_i|\bm{\beta, \gamma}).
	\label{eq:comp_alpha}
\end{equation}
The complexity regularizations for $\bm{\beta, \gamma}$ are similar to the Eq. (\ref{eq:comp_alpha}). The regularization $\mathcal{L}_{comp}$ is the weighted summation of $\mathcal{L}_{\alpha}$, $\mathcal{L}_{\beta}$ and $\mathcal{L}_{\gamma}$. With the regularization $\mathcal{L}_{comp}$ for regularizing the model complexity, our final regularization for searching is presented as follows:
\begin{align}
	\label{eq:loss_search}
	\mathcal{L}_{search}&=\mathcal{L}_{dp}+\lambda\mathcal{L}_{comp} \\\notag
	&=\mathcal{L}_{dp}+\lambda(\lambda_{\alpha}\mathcal{L}_{\alpha}+\lambda_{\beta}\mathcal{L}_{\beta}+\lambda_{\gamma}\mathcal{L}_{\gamma}),
\end{align}
where the $\lambda$ and $\lambda_{\alpha}, \lambda_{\beta}, \lambda_{\gamma}$ are the weights for complexity regularizations. We set $\lambda_{\alpha}=0.27, \lambda_{\beta}=0.27$ and $\lambda_{\gamma}=0.46$ based on the total time of each complexity regularization in Eq. (\ref{eq:comp_alpha}). Notably, although the complexity regularization $\mathcal{L}_{\alpha}$ is based on the inference time, it is still feasible to base on the parameter count or FLOPs.

\begin{algorithm}[tb]
	\caption{The overall pipeline.}
	\label{alg:pipeline}
	\begin{algorithmic}[1]
		\Require
		The search space $\mathcal{A}$, the training data $\mathcal{D}_{tr}$ divided into $\mathcal{D}_{tr}^w$ and $\mathcal{D}_{tr}^{archi}$, the pretrained model $\Omega$ consisting of $\Omega_{0}, \Omega_{1}, \Omega_{2}$, searchable architecture $S_{\{\bm{\alpha, \beta, \gamma, \delta, \bm{w}}\}}$ consisting of $S_{0}, S_{1}, S_{2}$.  
		\Ensure
		Trained denoising architecture $S_{\{\bm{\alpha^*, \beta^*, \gamma^*, \delta^*, w^*}\}}$.
		\State // Search step
		
		\For{$S_{i}$ in [$S_{0}, S_{1}, S_{2}$]}
		
		\State \parbox[t]{\dimexpr\linewidth-\algorithmicindent}{Initialize architecture weights $\{\bm{\alpha, \beta, \gamma, \delta}\}$ and operator parameters $\bm{w}$.}
		\While{not converge}
		
		\State \parbox[t]{\dimexpr\linewidth-\algorithmicindent-\algorithmicindent}{Update architecture weights $\{\bm{\alpha, \beta, \gamma, \delta}\}$ on $\mathcal{D}_{tr}^{archi}$ by descending $\nabla_{\{\bm{\alpha, \beta, \gamma, \delta}\}}\mathcal{L}_{search}$.}
		
		\State \parbox[t]{\dimexpr\linewidth-\algorithmicindent-\algorithmicindent}{ Update operator parameters $\bm{w}$ on $\mathcal{D}_{tr}^w$ by descending $\nabla_{\bm{w}}\mathcal{L}_{dp}$.}
		
		\EndWhile
		
		\State \parbox[t]{\dimexpr\linewidth-\algorithmicindent}{Derive the searched part $S_{i}$ based on $\{\bm{\alpha, \beta, \gamma, \delta}\}$.}
		
		\EndFor
		
		\State Derive the complete searched architecture $S_{\{\bm{\alpha^*, \beta^*, \gamma^*, \delta^*}\}}$
		
		\State // Train step
		
		\State Initialize operator parameters $\bm{w}$.
		
		\While{not converge}
		
		\State \parbox[t]{\dimexpr\linewidth-\algorithmicindent}{Update operator parameters $\bm{w}$ on $\mathcal{D}_{tr}$ by descending $\nabla_{\bm{w}}\mathcal{L}_{train}$.}
		
		\EndWhile
		
		\\
		\Return the final trained architecture $S_{\{\bm{\alpha^*, \beta^*, \gamma^*, \delta^*, w^*}\}}$.
	\end{algorithmic}
\end{algorithm}

\subsection{The overall pipeline}
\label{sec:overall}
Our pipeline is divided into $2$ steps. The first step is the search step, to find the optimal architecture weights $\{\bm{\alpha, \beta, \gamma, \delta}\}$. Once we obtain the architecture weights, we start the second step, decoding and training the denoising architecture. Significantly, in the train step, we only adopt the architecture weights $\{\bm{\alpha, \beta, \gamma, \delta}\}$ from the search step and abandon the parameters in operators $O_i({\boldsymbol x})$, which is common in most NAS methods, called train from scratch.

In the search step, we utilize the loss Eq. (\ref{eq:loss_search}) for searching. If we want to search an efficient architecture, we set $\lambda=0.001$. Otherwise, we set $\lambda=0$ to search the best-performing architecture regardless of complexity. In the train step, we train our searched architecture based on the loss that is represented as follows:
\begin{equation}
	\mathcal{L}_{train} = \mathcal{L}_{dp} + {\Vert I_{pred}-I_{gt} \Vert_1},
	\label{eq:loss_train}
\end{equation}
where $I_{pred}, I_{gt}$ are the prediction and ground-truth images, respectively. 
We optimize our searched architecture with $\mathcal{L}_{train}$ for the first few epochs and then discard the loss $\mathcal{L}_{dp}$ for further training. The algorithm of our overall pipeline is provided in Algorithm \ref{alg:pipeline}.


\section{Experiments}
We first introduce the datasets and implementation details. Next, we compare our method with NAS-based denoising methods and state-of-the-art methods. Then, we evaluate the components of our method. At last, we discuss our experimental results to draw some directions for further works. 
\subsection{Datasets}
\begin{figure*}[tp]
	\centering
	\captionsetup[subfloat]{labelsep=none,format=plain,labelformat=empty,font=footnotesize}
	\begin{minipage}{.35\textwidth}
		\subfloat[CBSD68: 3072 (Case 2)]{
		  \includegraphics[width=.95\textwidth]{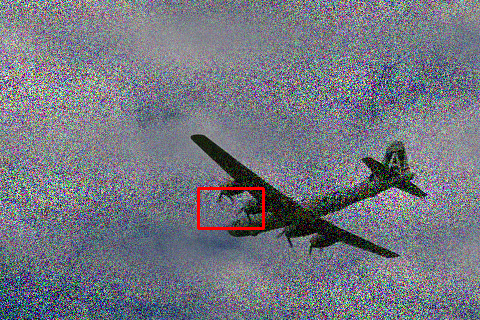}}
	\end{minipage}
	\begin{minipage}{.15\textwidth}
		\subfloat[Noisy: 14.90 dB]{
		  \includegraphics[width=.95\textwidth]{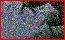}}\\
		\subfloat[NBNet: 36.54 dB]{
		  \includegraphics[width=.95\textwidth]{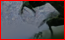}}
	\end{minipage}
	\begin{minipage}{.15\textwidth}
		\subfloat[DnCNN: 36.64 dB]{
		  \includegraphics[width=.95\textwidth]{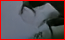}}\\
		\subfloat[Restormer: 37.21 dB]{
		  \includegraphics[width=.95\textwidth]{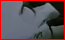}}
	\end{minipage}
	\begin{minipage}{.15\textwidth}
		\subfloat[VDN: 36.98 dB]{
		  \includegraphics[width=.95\textwidth]{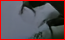}}\\
		\subfloat[Ours: 37.78 dB]{
		  \includegraphics[width=.95\textwidth]{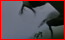}}
	\end{minipage}
	\begin{minipage}{.15\textwidth}
		\subfloat[RDN: 37.19 dB]{
		  \includegraphics[width=.95\textwidth]{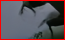}}\\
		\subfloat[GT]{
		  \includegraphics[width=.95\textwidth]{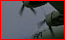}}
	\end{minipage}
	\caption{Visual quality comparison. The input image was cropped from the CBSD68 dataset and contaminated by the non-i.i.d noise in `Case 2'.}
	\label{fig:niid}
\end{figure*}

\begin{figure*}[tp]
	\centering
	\captionsetup[subfloat]{labelsep=none,format=plain,labelformat=empty,font=footnotesize}
	\begin{minipage}{.35\textwidth}
		\subfloat[LIVE1: buildings ($\sigma=50$)]{
		  \includegraphics[width=.95\textwidth]{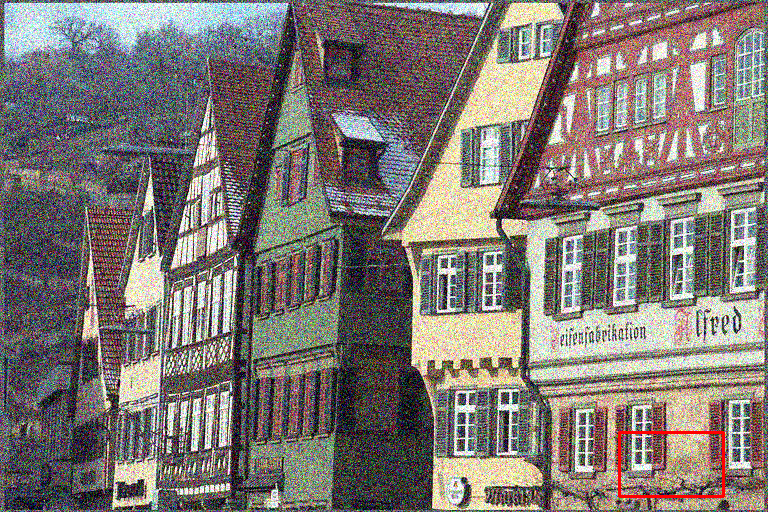}}
	\end{minipage}
	\begin{minipage}{.15\textwidth}
		\subfloat[Noisy: 14.88 dB]{
		  \includegraphics[width=.95\textwidth]{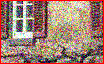}}\\
		\subfloat[NBNet: 27.04 dB]{
		  \includegraphics[width=.95\textwidth]{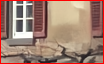}}
	\end{minipage}
	\begin{minipage}{.15\textwidth}
		\subfloat[DnCNN: 26.51 dB]{
		  \includegraphics[width=.95\textwidth]{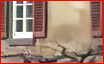}}\\
		\subfloat[Restormer: 27.99 dB]{
		  \includegraphics[width=.95\textwidth]{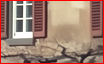}}
	\end{minipage}
	\begin{minipage}{.15\textwidth}
		\subfloat[VDN: 26.82 dB]{
		  \includegraphics[width=.95\textwidth]{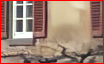}}\\
		\subfloat[Ours: 28.69 dB]{
		  \includegraphics[width=.95\textwidth]{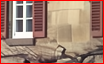}}
	\end{minipage}
	\begin{minipage}{.15\textwidth}
		\subfloat[MIRNet: 27.39 dB]{
		  \includegraphics[width=.95\textwidth]{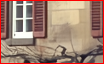}}\\
		\subfloat[GT]{
		  \includegraphics[width=.95\textwidth]{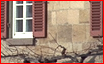}}
	\end{minipage}
	\caption{Visual quality comparison. The input image was cropped from the LIVE1 dataset and contaminated by AWGN with a large noise intensity $\sigma = 50$.}
	\label{fig:awgn}
\end{figure*}

For synthetic image denoising, we apply our method to two types of noises, the i.i.d noise, i.e., the additive white Gaussian noise (AWGN), and the non-i.i.d noise, introduced by VDN \cite{yue2019variational}. The differences between the two types of noises can be observed in Figure \ref{fig:niid} and Figure \ref{fig:awgn}. The non-i.i.d noise is modeled as a non-i.i.d and pixel-wise Gaussian distribution that is more similar to the complicated real-world noise, which is simulated as:
\begin{equation}
	n=n^1\odot M, n^1_{ij} \sim \mathcal{N}(0,1),
\end{equation}  	    	   
where $M$ is a spatially variant map. There are 4 kinds of maps generated. One is for training, and the others are for testing, which are cataloged as Case 1, Case 2, and Case 3. The corresponding clean images for training are sampled from three datasets, including 432 images from BSD500 \cite{arbelaez2010contour}, 400 images from the validation of ImageNet \cite{krizhevsky2012imagenet} and 4744 images from the Waterloo dataset \cite{ma2016waterloo}. We choose CBSD68, Set5, and LIVE1 as the source of test datasets.

For real-world image denoising, we evaluate our method in the Smartphone Image Denoising Dataset (SIDD) \cite{SIDD_2018_CVPR, Abdelhamed_2019_CVPR_Workshops} and the Darmstadt Noise Dataset (DND) \cite{plotz2017benchmarking}. The medium version of SIDD consists of 320 pairs adopted for searching and training, while the official SIDD Validation Data and DND are used for evaluation. The DND does not provide pair-wise images.

\subsection{Inplementation Details}
\paragraph{Search step} For the synthetic datasets, there are 5576 image pairs. We randomly divide the images into two parts. A total of 3000 image pairs are utilized to update the parameters $\bm{w}$ in operators, and 2576 pairs are used to update the architecture weights $\{\bm{\alpha, \beta, \gamma, \delta}\}$. All the images are cropped to $128\times128$ patches. We search the architecture with a batch size of 16 for 150 epochs. The ADAM optimizers are adopted to alternately optimize the operator parameters and the architecture weights with fixed learning rates of $2e$-$4$ and $1e$-$4$, respectively. We utilize a pretrained MIRNet \cite{zamir2020learning} as the denoising prior in Eq. \ref{eq:dp}. The MIRNet has $3$ parts that have the same components. For the real-world noisy images, 160 image pairs of the SIDD-Medium dataset are for operator parameters and the remaining 160 pairs are for architecture weights. The other settings are the same as the synthetic datasets.

\paragraph {Train step} Once we obtain the architecture weights $\{\bm{\alpha, \beta, \gamma, \delta}\}$, the way for training the searched architecture is similar to other deep learning denoising methods. We train the searched architecture with $128\times128$ patches, and the batch size is 32. As described in Section \ref{sec:overall}, we first apply the loss in Eq. (\ref{eq:loss_train}) for 60 epochs, then optimize the architecture only by the L1 loss for 240 epochs. The optimizer is the ADAM optimizer, and the learning rate decays from $2e$-$4$ to $1e$-$6$ with the cosine annealing strategy. 

\begin{table*}[htp]
	\centering
	\caption{Comparisons with NAS-based methods of image denoising. $\times 3$ represents that we parallelly search for the optimal architecture in $3$ parts.}
	\begin{tabular}{@{}cccccccc@{}}
	\toprule
	Methods & \# Candidates & GPU         & Hours & Params (M) & FLOPs (G) & PSNR           & SSIM           \\ \midrule
	CLEARER~\cite{gou2020clearer} & 4096                    & 1 1080 Ti   & 20    & 9.17        & 92.30     & 38.70          & 0.950          \\
	HiNAS~\cite{zhang2021memory}   & $\approx1$e$42$   		 & 1 1080 Ti   & 34    & 1.53        & 100.20    & 38.56          & 0.949          \\
	Ours    & $\approx\textbf{1}$\textbf{e}$\textbf{1800}$         & $4\times3$ 1080 Ti & $48\times3$  & 8.01        & 98.51     & \textbf{40.12} & \textbf{0.960} \\ \bottomrule
	\end{tabular}
	\label{tab:nas-based}
\end{table*}

\begin{table*}[htp]
	\centering
	\caption{The average PSNR/SSIM on real-world image denoising: SIDD and DND datasets. The FLOPs were computed on image size 256×256. The inference time was computed on the SIDD datasets that consist of 1280 images on size 256×256.}
	\begin{tabular}{ccccccccccccc}
		\toprule
		 & \begin{tabular}[c]{@{}c@{}}DnCNN\\ \cite{zhang2017beyond}\end{tabular} & \begin{tabular}[c]{@{}c@{}}RIDNet\\ \cite{anwar2019real}\end{tabular} & \begin{tabular}[c]{@{}c@{}}DANet\\ \cite{yue2020dual}\end{tabular} & \begin{tabular}[c]{@{}c@{}}VDN\\ \cite{yue2019variational}\end{tabular} & \begin{tabular}[c]{@{}c@{}}MIRNet\\ \cite{zamir2020learning}\end{tabular} & \begin{tabular}[c]{@{}c@{}}DeamNet\\ \cite{ren2021adaptive}\end{tabular} & \begin{tabular}[c]{@{}c@{}}InvDN\\ \cite{liu2021invertible}\end{tabular} & \begin{tabular}[c]{@{}c@{}}NBNet\\ \cite{cheng2021nbnet}\end{tabular} & \begin{tabular}[c]{@{}c@{}}VDIR+\\ \cite{soh2022variational}\end{tabular} & \begin{tabular}[c]{@{}c@{}}Restormer\\ \cite{zamir2022restormer}\end{tabular} & \begin{tabular}[c]{@{}c@{}}Uformer\\ \cite{wang2022uformer}\end{tabular} & Ours \\ \hline
		\multirow{2}{*}{SIDD} & 23.66 & 38.71 & 39.47 & 39.28 & 39.72 & 39.40 & 39.28 & 39.68 & 39.33 & 39.93 & 39.81 & \textbf{40.12} \\ \cline{2-13} 
		 & 0.583 & 0.951 & 0.957 & 0.956 & 0.959 & 0.957 & 0.955 & 0.958 & 0.956 & \textbf{0.960} & 0.959 & \textbf{0.960} \\ \hline
		\multirow{2}{*}{DND} & 32.43 & 39.26 & 39.58 & 39.38 & 39.99 & 39.63 & 39.57 & 39.89 & 39.69 & 40.03 & 39.98 & \textbf{40.18} \\ \cline{2-13} 
		 & 0.790 & 0.953 & 0.955 & 0.952 & 0.956 & 0.953 & 0.952 & 0.955 & 0.953 & 0.956 & 0.955 & \textbf{0.961} \\ \hline
		Params (M) & 0.56 & 1.50 & 9.15 & 7.82 & 31.79 & 2.23 & 2.64 & 13.31 & 2.20 & 26.10 & 50.88 & 8.01 \\ \hline
		FLOPs (G) & 36.67 & 98.13 & 14.85 & 49.50 & 787.04 & 146.18 & 406.33 & 88.79 & - & 140.99 & 89.46 & 98.51 \\ \hline
		Inference time (s) & 11.10 & 27.04 & 5.64 & 10.50 & 160.53 & 51.12 & 49.11 & 30.51 & - & 111.57 & 80.15 & 70.33 \\ 
		\bottomrule
		\end{tabular}
	\label{tab:real}
\end{table*}

\begin{figure*}[tp]
	\centering
	\captionsetup[subfloat]{labelsep=none,format=plain,labelformat=empty,font=footnotesize}
	\subfloat[Noisy: 18.16 dB]{
	\begin{minipage}[c]{0.18\textwidth}
			\centering
			\begin{overpic}[width=1\textwidth]{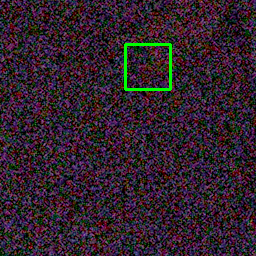}
				\put(50, 0){\includegraphics[scale=1]{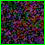}}
			\end{overpic}
		\end{minipage}}
	\subfloat[RIDNet: 29.83 dB]{
	\begin{minipage}[c]{0.18\textwidth}
			\centering
			\begin{overpic}[width=1\textwidth]{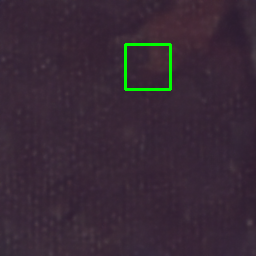}
				\put(50, 0){\includegraphics[scale=1]{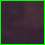}}
			\end{overpic}
		\end{minipage}}
	\subfloat[VDN: 30.31 dB]{
	\begin{minipage}[c]{0.18\textwidth}
			\centering
			\begin{overpic}[width=1\textwidth]{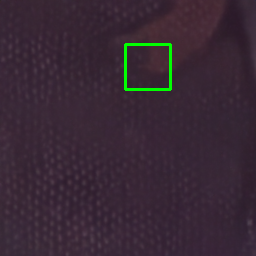}
				\put(50, 0){\includegraphics[scale=1]{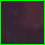}}
			\end{overpic}
		\end{minipage}}
	\subfloat[MIRNet: 31.36 dB]{
	\begin{minipage}[c]{0.18\textwidth}
			\centering
			\begin{overpic}[width=1\textwidth]{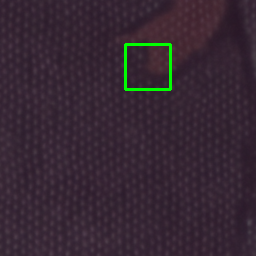}
				\put(50, 0){\includegraphics[scale=1]{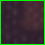}}
			\end{overpic}
		\end{minipage}}
	\subfloat[InvDN: 29.89 dB]{
	\begin{minipage}[c]{0.18\textwidth}
			\centering
			\begin{overpic}[width=1\textwidth]{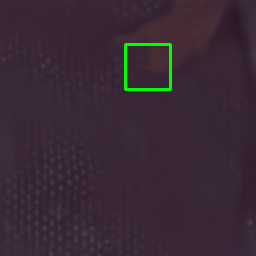}
				\put(50, 0){\includegraphics[scale=1]{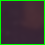}}
			\end{overpic}
		\end{minipage}}
	
	\subfloat[DeamNet: 30.25 dB]{
	\begin{minipage}[c]{0.18\textwidth}
		\centering
		\begin{overpic}[width=1\textwidth]{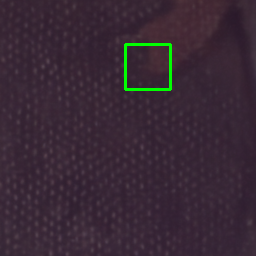}
			\put(50, 0){\includegraphics[scale=1]{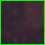}}
		\end{overpic}
	\end{minipage}}
	\subfloat[NBNet: 30.18 dB]{
	\begin{minipage}[c]{0.18\textwidth}
		\centering
		\begin{overpic}[width=1\textwidth]{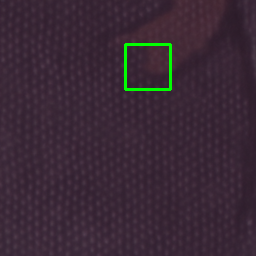}
			\put(50, 0){\includegraphics[scale=1]{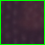}}
		\end{overpic}
	\end{minipage}}
	\subfloat[Uformer: 31.37 dB]{
	\begin{minipage}[c]{0.18\textwidth}
		\centering
		\begin{overpic}[width=1\textwidth]{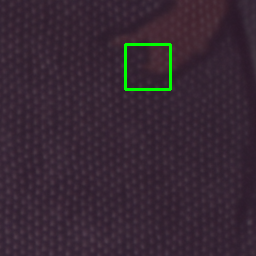}
			\put(50, 0){\includegraphics[scale=1]{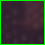}}
		\end{overpic}
	\end{minipage}}
	\subfloat[Restormer: 31.57 dB]{
	\begin{minipage}[c]{0.18\textwidth}
			\centering
			\begin{overpic}[width=1\textwidth]{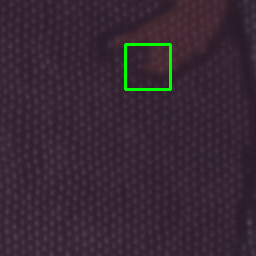}
				\put(50, 0){\includegraphics[scale=1]{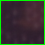}}
			\end{overpic}
		\end{minipage}}
	\subfloat[Ours: 31.82 dB]{
	\begin{minipage}[c]{0.18\textwidth}
			\centering
			\begin{overpic}[width=1\textwidth]{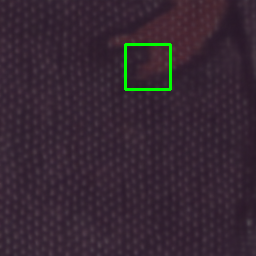}
				\put(50, 0){\includegraphics[scale=1]{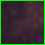}}
			\end{overpic}
		\end{minipage}}
		\caption{Visual quality comparison. The input image was cropped from the SIDD dataset.}
		\label{fig:sidd}
\end{figure*}
\subsection{Comparisons with other NAS-based denoising methods}
To evaluate our method, we compare 2 NAS-based denoising methods on several metrics on the SIDD dataset. As shown in Table \ref{tab:nas-based}, our searched architecture is superior to HiNAS \cite{zhang2020memory, zhang2021memory} and CLEARER \cite{gou2020clearer} by a large margin. It surpasses them by 1.50 dB with similar model complexity. This is largely due to the experience we learned from existing denoising methods, i.e., the specific denoising search space and the denoising priors. Meanwhile, as illustrated in Table \ref{tab:strategies}, their search strategies do not allow such a giant search space like ours, i.e., approximately $\textbf{1}$\textbf{e}$\textbf{1800}$.

\begin{figure*}[tp]
	\centering
	\captionsetup[subfloat]{labelsep=none,format=plain,labelformat=empty,font=footnotesize}
	\subfloat[Noisy: 18.16 dB]{
	\begin{minipage}[c]{0.18\textwidth}
			\centering
			\begin{overpic}[width=1\textwidth]{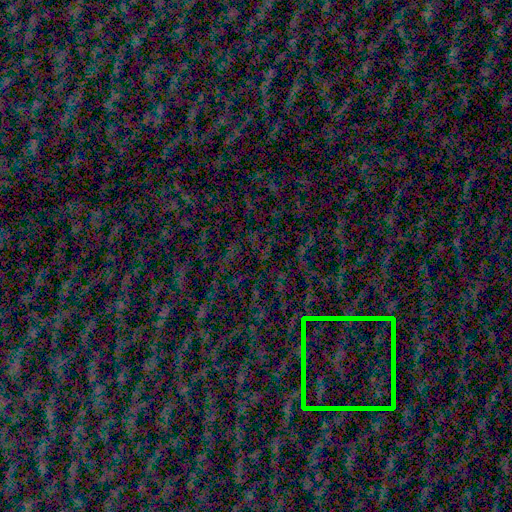}
				\put(0, 0){\includegraphics[scale=0.5]{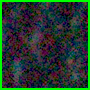}}
			\end{overpic}
		\end{minipage}}
	\subfloat[InvDN: 32.84 dB]{
	\begin{minipage}[c]{0.18\textwidth}
			\centering
			\begin{overpic}[width=1\textwidth]{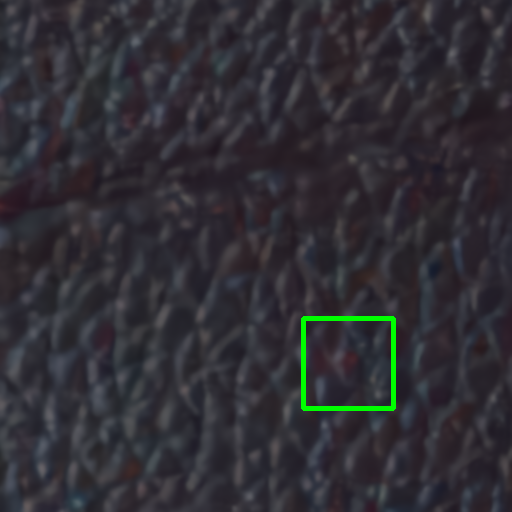}
				\put(0, 0){\includegraphics[scale=0.5]{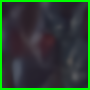}}
			\end{overpic}
		\end{minipage}}
	\subfloat[VDIR: 32.26 dB]{
			\begin{minipage}[c]{0.18\textwidth}
					\centering
					\begin{overpic}[width=1\textwidth]{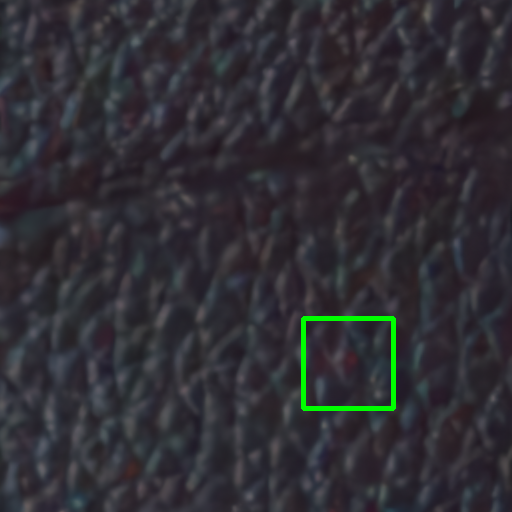}
						\put(0, 0){\includegraphics[scale=0.5]{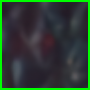}}
					\end{overpic}
				\end{minipage}}
	\subfloat[Restormer: 33.04 dB]{
	\begin{minipage}[c]{0.18\textwidth}
			\centering
			\begin{overpic}[width=1\textwidth]{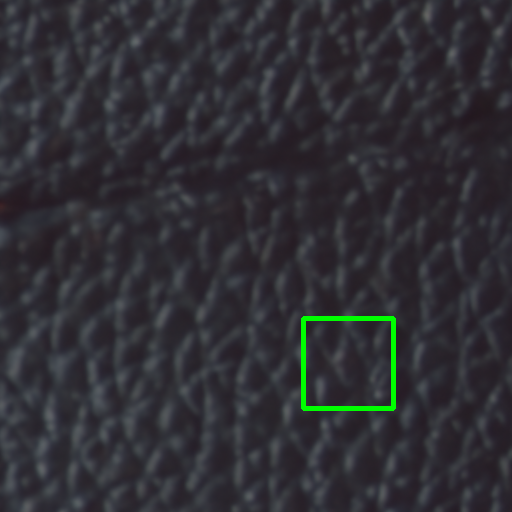}
				\put(0, 0){\includegraphics[scale=0.5]{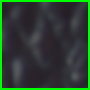}}
			\end{overpic}
		\end{minipage}}
	\subfloat[Ours: 34.42 dB]{
	\begin{minipage}[c]{0.18\textwidth}
			\centering
			\begin{overpic}[width=1\textwidth]{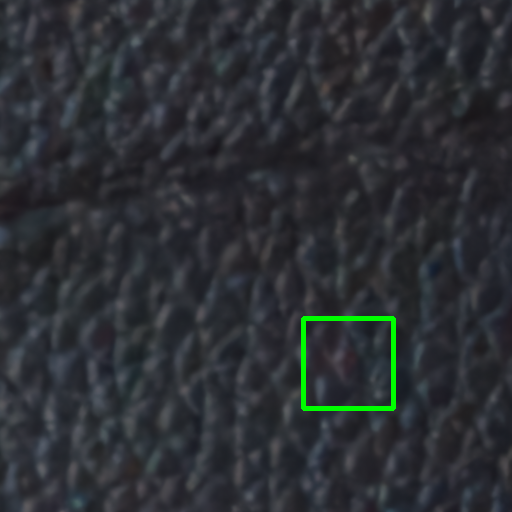}
				\put(0, 0){\includegraphics[scale=0.5]{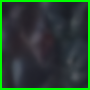}}
			\end{overpic}
		\end{minipage}}
	\caption{Visual quality comparison. The input image was cropped from the DND dataset.}
	\label{fig:dnd}
\end{figure*}

\begin{figure*}
	\centering
	\includegraphics[width=\linewidth]{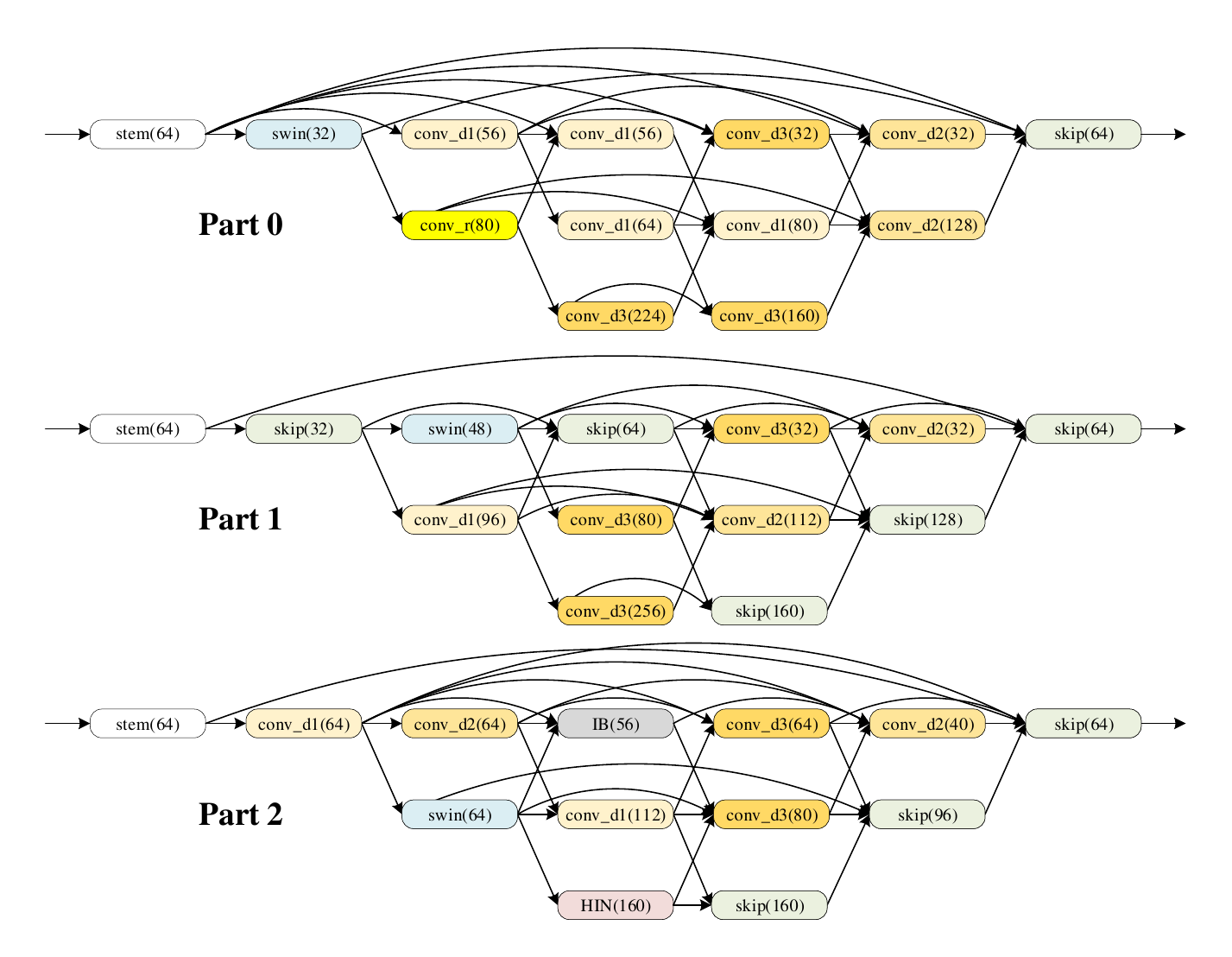}
	\caption{The searched architecture. The yellow box is our searched cells. For example, the `swin(32)' indicates the selected operator is `swin' with the output dimension 32. The `stem(64)' is the vanilla convolution with dimension 64. In particular, `skip(64)' represents the features are taken from index 0 to index 63, and this is mentioned in Section \ref{sec:Coarse-to-fine}.}
	\label{fig:searched}
\end{figure*}

\subsection{Comparisons with state-of-the-art methods}
We compare our method with $14$ SOTA methods on all datasets, including $1$ traditional method and $13$ deep learning-based manual methods. Table \ref{tab:real} illustrates the comparisons on the real-world datasets, and the searched architecture is shown in Figure \ref{fig:searched}. Our method achieves PSNR gain against Restormer \cite{zamir2022restormer} by about 0.2 dB on SIDD and 0.15 dB on DND while requires fewer FLOPs, i.e., $40G$ less than that in Restormer. Also, the parameter count of our searched architecture is only $\frac{1}{6}$ of Uformer \cite{wang2022uformer}. When compared with VDIR+ \cite{soh2022variational}, the improvement is up to 0.8 dB and 0.5 dB, respectively. We also visualize some results of each dataset in Figure \ref{fig:sidd} and Figure \ref{fig:dnd}. On the SIDD dataset, compared with other methods, we retain more textures, especially the `yellow pattern'. On the DND dataset, we remove the noise while retaining the color and the structure. 

The comparisons in synthetic datasets are shown in Table \ref{tab:syn}, and we achieve the best performance in all cases. The visual results can be observed in Figure \ref{fig:niid} and Figure \ref{fig:awgn}. We introduce fewer artifacts while removing the noises.

\begin{table*}[tp]
	\centering
	\caption{The average PSNR on synthetic datasets: CBSD68, LIVE1, and Set5.}
	\resizebox{0.9\textwidth}{!}{
	\begin{tabular}{lcccccccccc}
		\toprule
										&        & CBM3D~\cite{joshi2009image} & DnCNN~\cite{zhang2017beyond} & FFDNet~\cite{zhang2018ffdnet} & VDN~\cite{yue2019variational}   & MIRNet~\cite{zamir2020learning} & RDN~\cite{zhang2020residual}   & NBNet~\cite{cheng2021nbnet} & Restormer~\cite{zamir2022restormer} & Ours  \\ \hline
		\multirow{3}{*}{Case 1}      & CBSD68 & 26.51 & 28.73 & 28.78  & 29.02 & 29.07  & 29.01 & 29.16 & 29.25     & \textbf{29.36} \\ \cline{2-11} 
										& LIVE1  & 26.58 & 28.81 & 28.99  & 29.22 & 29.33  & 29.28 & 29.40 & 29.54     & \textbf{29.67} \\ \cline{2-11} 
										& Set5   & 27.76 & 29.85 & 30.16  & 30.39 & 30.46  & 30.43 & 30.59 & 30.68     & \textbf{30.82} \\ \hline
		\multirow{3}{*}{Case 2}      & CBSD68 & 25.28 & 28.15 & 28.43  & 28.67 & 28.72  & 28.70 & 28.76 & 28.88     & \textbf{28.99} \\ \cline{2-11} 
										& LIVE1  & 25.18 & 28.18 & 28.58  & 28.82 & 28.94  & 28.88 & 29.01 & 29.12     & \textbf{29.26} \\ \cline{2-11} 
										& Set5   & 26.34 & 29.04 & 29.60  & 29.80 & 29.93  & 29.81 & 29.88 & 29.90     & \textbf{30.12} \\ \hline
		\multirow{3}{*}{Case 3}      & CBSD68 & 26.44 & 28.11 & 28.22  & 28.46 & 28.50  & 28.49 & 28.59 & 28.66     & \textbf{28.79} \\ \cline{2-11} 
										& LIVE1  & 26.50 & 28.17 & 28.39  & 28.65 & 28.74  & 28.69 & 28.82 & 28.89     & \textbf{29.01} \\ \cline{2-11} 
										& Set5   & 27.88 & 29.13 & 29.54  & 29.74 & 29.83  & 29.79 & 29.89 & 29.93     & \textbf{30.13} \\ \hline
		\multirow{3}{*}{$\sigma=15$} & CBSD68 & 32.67 & 33.87 & 33.85  & 33.90 & 34.09  & 34.00 & 34.15 & 34.39     & \textbf{34.49} \\ \cline{2-11} 
										& LIVE1  & 32.85 & 33.72 & 33.96  & 33.94 & 34.20  & 34.03 & 34.25 & 34.59     & \textbf{34.71} \\ \cline{2-11} 
										& Set5   & 33.42 & 34.04 & 34.30  & 34.34 & 34.59  & 34.51 & 34.64 & 35.02     & \textbf{35.07} \\ \hline
		\multirow{3}{*}{$\sigma=25$} & CBSD68 & 29.83 & 31.22 & 31.21  & 31.35 & 31.48  & 31.41 & 31.54 & 31.78     & \textbf{31.89} \\ \cline{2-11} 
										& LIVE1  & 30.05 & 31.23 & 31.37  & 31.50 & 31.70  & 31.60 & 31.73 & 32.09     & \textbf{32.18} \\ \cline{2-11} 
										& Set5   & 30.92 & 31.88 & 32.10  & 32.24 & 32.44  & 32.33 & 32.51 & 32.90     & \textbf{32.92} \\ \hline
		\multirow{3}{*}{$\sigma=50$} & CBSD68 & 26.81 & 27.91 & 27.95  & 28.19 & 28.25  & 28.22 & 28.35 & 28.60     & \textbf{28.71} \\ \cline{2-11} 
										& LIVE1  & 26.98 & 27.95 & 28.10  & 28.36 & 28.48  & 28.39 & 28.55 & 28.92     & \textbf{29.04} \\ \cline{2-11} 
										& Set5   & 28.16 & 28.95 & 29.25  & 29.47 & 29.52  & 29.50 & 29.70 & 30.02     & \textbf{30.10} \\ 
	\bottomrule
	\end{tabular}}
	\label{tab:syn}
\end{table*}

\begin{figure*}[htp]
	\centering
	\subfloat[]{\includegraphics[width=0.45\linewidth]{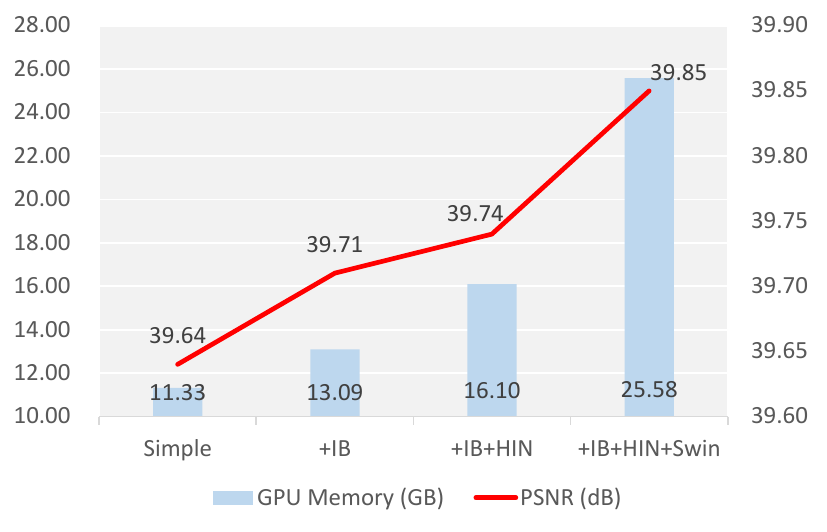}%
		\label{fig:ops}}
	\subfloat[]{\includegraphics[width=0.45\linewidth]{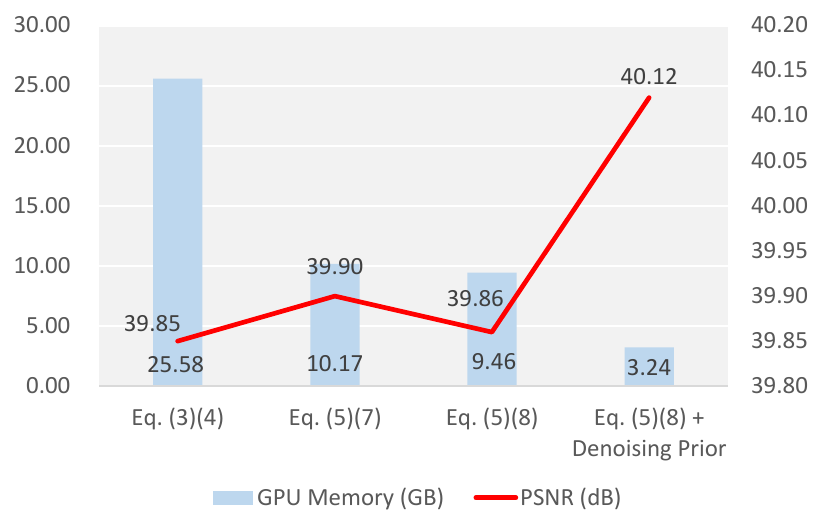}%
		\label{fig:strategy}}
	\caption{(a) The denoising performance is improved when specific denoising operators are injected into the space as shown in red line, and the GPU memory usage proliferates due to the search strategy (Eqs. (\ref{eq:darts}) (\ref{eq:gra_darts})) widely adopted in NAS-based methods as shown in blue bins. (b) Our modification gradually reduces the GPU memory with the such a giant search space.}
	\label{fig:blocks_stra}
\end{figure*}

\subsection{Ablation study}
In our ablation study, we first verify the necessity and the difficulty of introducing specific denoising operators. Then, we study the advantages brought by each level search space for image denoising. Finally, we evaluate the effects of our two regularizations. 

\textbf{Influence of injected operators and different search strategies.} As shown in Figure \ref{fig:ops}, with the injection of more specific denoising operators, the architecture performs better. This is consistent with the observation in Figure \ref{fig:feat_mean_std}, as the different specific denoising operators focus on diverse perspectives of image denoising. Meanwhile, the increasing search space in cell-level search may give rise to the overflow of GPU memory in existing NAS-based methods for denoising. Hence, we modify the search strategies while introducing these necessary operators. As shown in Figure \ref{fig:strategy}, the GPU memory is relaxed, and this phenomenon is consistent with Table \ref{tab:strategies}.

\begin{table}[tp]
	\centering
	\caption{Ablation study on the effect of each-level search on the SIDD dataset.}
	\setlength{\tabcolsep}{1.1mm}{
	\begin{tabular}{@{}cccccc@{}}
		\toprule
		\multicolumn{2}{c}{Network-level}                     & \multirow{2}{*}{Cell-level} & \multirow{2}{*}{Kernel-level} & \multirow{2}{*}{Params (M)} & \multirow{2}{*}{PSNR} \\ \cline{1-2}
		Resolution                & Connection                &                             &                               &                              &                       \\ \hline
		\XSolidBrush    & \Checkmark & \Checkmark   & \Checkmark     & 7.60                         & 39.47                 \\
		\Checkmark & \XSolidBrush    & \Checkmark   & \Checkmark     & 8.33                         & 39.70                 \\
		\Checkmark & \Checkmark & \XSolidBrush      & \Checkmark     & 8.26                         & 39.81                 \\
		\Checkmark & \Checkmark & \Checkmark   & \XSolidBrush        & 8.89                         & 39.90                 \\
		\Checkmark & \Checkmark & \Checkmark   & \Checkmark     & 8.01                         & 40.12        \\\bottomrule
	\end{tabular}}
	\label{tab:each-level}
\end{table}

\begin{table}[tp]
	\caption{Ablation study on the effect of our denoising prior-based regularization on the SIDD dataset. `BS' is the batchsize on each GPU.}
	\setlength{\tabcolsep}{0.4mm}{
	\begin{tabular}{@{}cccccc@{}}
	\toprule
	\multirow{2}{*}{\makecell[c]{Denoising Prior-based\\Regularization}} & \multicolumn{4}{c}{Search Cost}                          & \multirow{2}{*}{PSNR} \\ \cmidrule(lr){2-5}
															& GPU    & BS & GPU-Memory & Hours &                       \\ \midrule
	\XSolidBrush                                            & 4 1080 Ti & 2                     & 9.46            & 237   & 39.86                 \\
	\Checkmark                                              & $4\times3$ 1080 Ti & 2                     & 3.24            & $48\times3$    & 40.12                 \\ \bottomrule
	\end{tabular}}
	\label{tab:prior_denoise}
\end{table}

\begin{table}[tp]
	\caption{Ablation study on the effect of different complexity regularization on the SIDD dataset.}
	\setlength{\tabcolsep}{1.1mm}{
	\begin{tabular}{@{}ccccc@{}}
	\toprule
	\makecell[c]{Complexity\\Regularization} & Params (M) & FLOPs (G) & Inference time (s) & PSNR  \\ \midrule
	Parameter-based           & 3.02        & 57.28     & 64.27              & 39.72 \\
	Flops-based               & 4.26        & 34.60     & 60.95              & 39.65 \\
	Inference time-based      & 4.51        & 43.32     & 60.59              & 39.66 \\ \bottomrule
	\end{tabular}}
	\label{tab:prior_complex}
\end{table}
\textbf{Influence of search space.} Our method introduces a coarse-to-fine search space inspired by the existing denoising methods. In Table \ref{tab:each-level}, we explore its effects. The `cross' symbol indicates that the architecture weights at this level are randomly initialized. All levels of search space boosts denoising. In addition, the performance drops the most with the lack of the multi-resolution search, as the operators receive incompatible features that may lose non-trivial information. 

\textbf{Influence of denoising prior-based regularization.} In Table \ref{tab:prior_denoise}, the denoising prior-based regularization gives feasibility to parallelly search for each part. As indicated in the GPU column, if the denoising prior-based regularization is not adopted, we have to search the complete architecture on $4$ 1080 Ti GPUs. After adding this regularization, the architecture is divided into $3$ parts and the optimization of each part is distributed on $4$ 1080 Ti GPUs. Therefore, we are able to search the optimal architecture on $4\times3$ 1080 Ti GPUs.
The search speed is significantly boosted due to the efficient regularization of search space for each part. Meanwhile, the denoising performance is also improved. This gain indicates that denoising prior-based regularization is beneficial for reducing the searching complexity and boosting the denoising performance. 

\textbf{Influence of different complexity regularization.} As mentioned in Section \ref{sec:regularization}, there are $3$ frequent metrics for measuring the model complexity, i.e., the parameter count, the FLOPs, and the inference time. These could be introduced to regularize the model complexity in our search. In Table \ref{tab:prior_complex}, we find that the inference time is not always consistent with the parameter count and the FLOPs. 
The model that optimized with the parameter-based regularization tends to have a smaller parameter count but at a slower speed. The model regularized by inference time has higher FLOPs than that by FLOPs-based regularization, but the former is faster.
This inconsistency could also be found in Table \ref{tab:real}. Although the FLOPs of DeamNet are similar to that of Restormer, the speed of DeamNet is much faster in inference. This is due to the optimizations for different operators, i.e., different operators may have similar parameter counts or FLOPs, but running time is different. This may be more obvious on edge devices. As a result, we recommend taking the inference time-based regularization to search for an efficient model.

\begin{figure*}[tp]
	\centering
	\subfloat[Models-Large]{\includegraphics[width=0.5\linewidth]{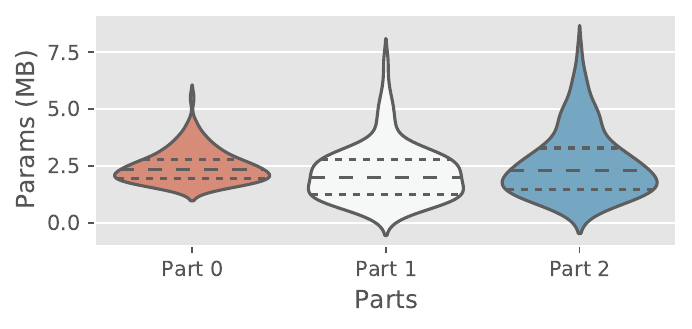}}%
	\subfloat[Models-Small]{\includegraphics[width=0.5\linewidth]{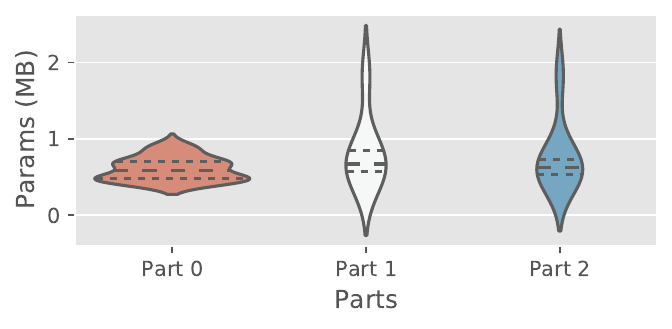}}%
	\hfil
	\subfloat[Models-Large]{\includegraphics[width=0.5\linewidth]{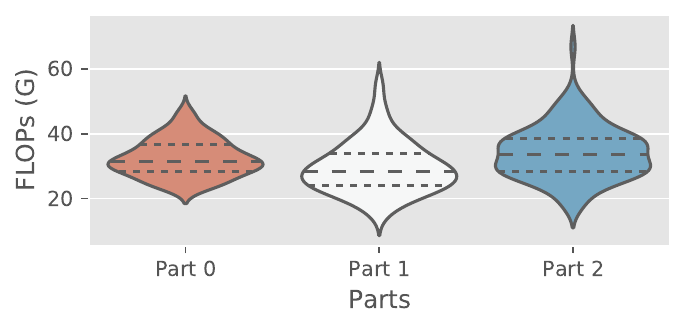}}%
	\subfloat[Models-Small]{\includegraphics[width=0.5\linewidth]{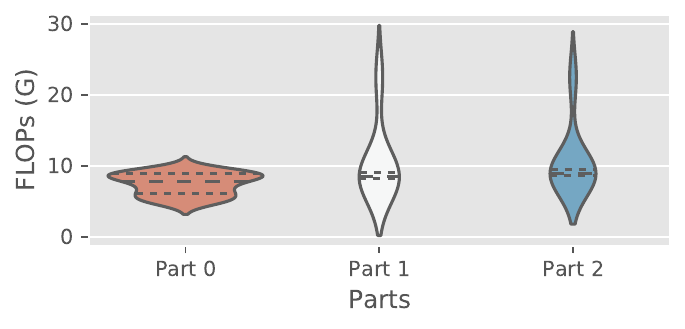}}%

	\caption{The distributions of model complexity of each part. The width of each part corresponds with the approximate frequency of the model under this FLOPs. The inner lines of the violins represent the quartiles. The `Models-Large' indicates the searched architectures with more complexity, while the `Models-Small' indicates those with less complexity.}
	\label{fig:big_small}
\end{figure*}

\begin{figure}[tp]
	\centering
	\includegraphics[width=\linewidth]{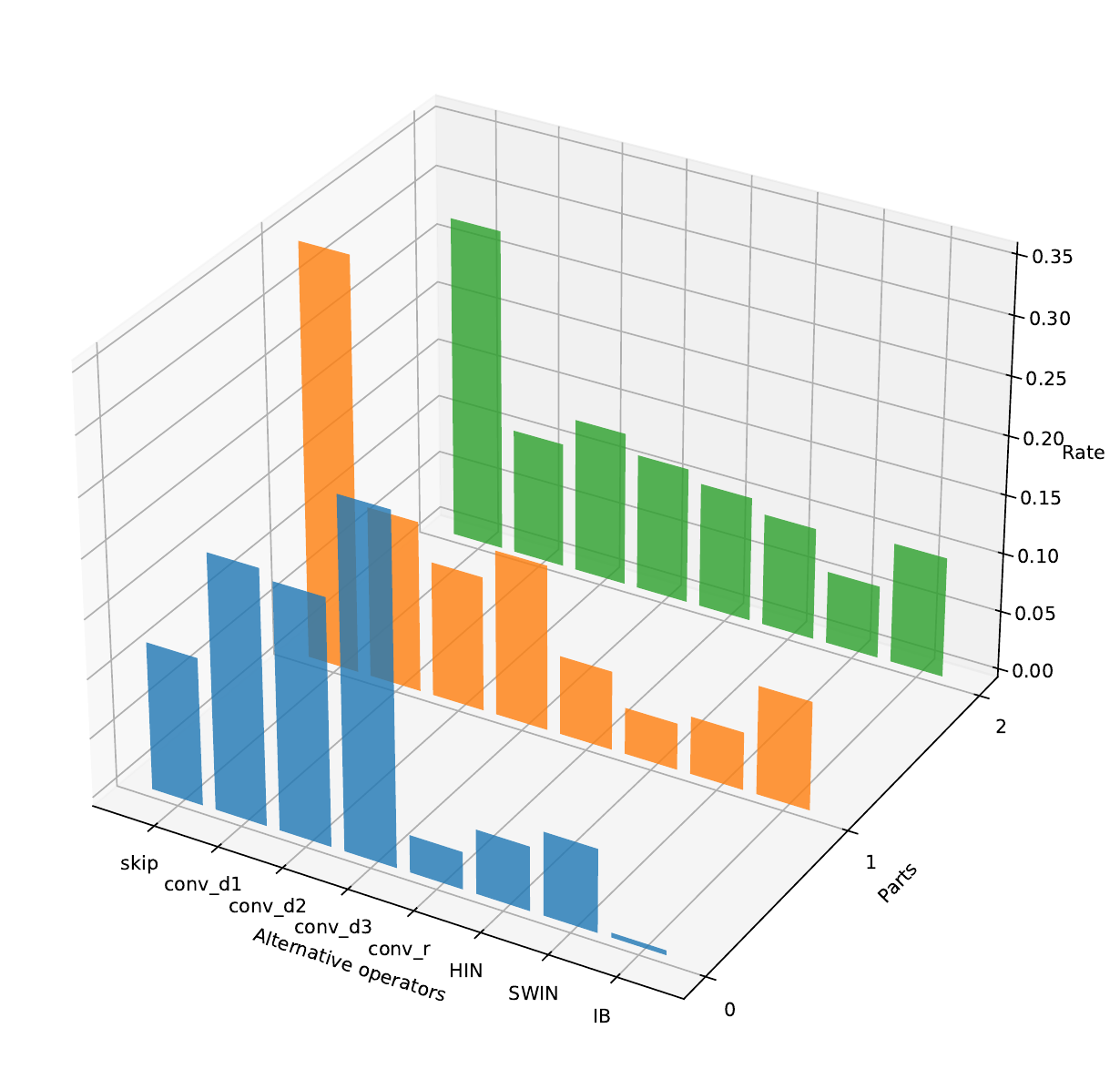}
	\caption{The rates of selected operators in `Part 0', `Part 1', and `Part 2'.}
	\label{fig:hist} 
\end{figure}

\begin{figure}[tp]
	\centering
	\subfloat[Part 0]{\includegraphics[width=0.3\linewidth]{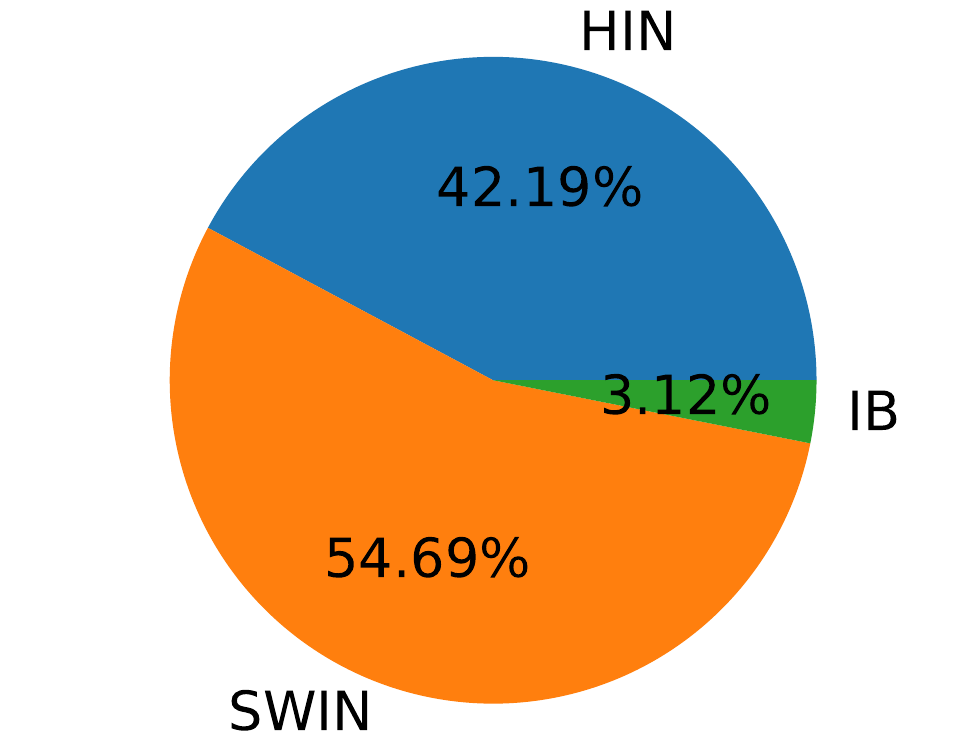}}%
	\subfloat[Part 1]{\includegraphics[width=0.3\linewidth]{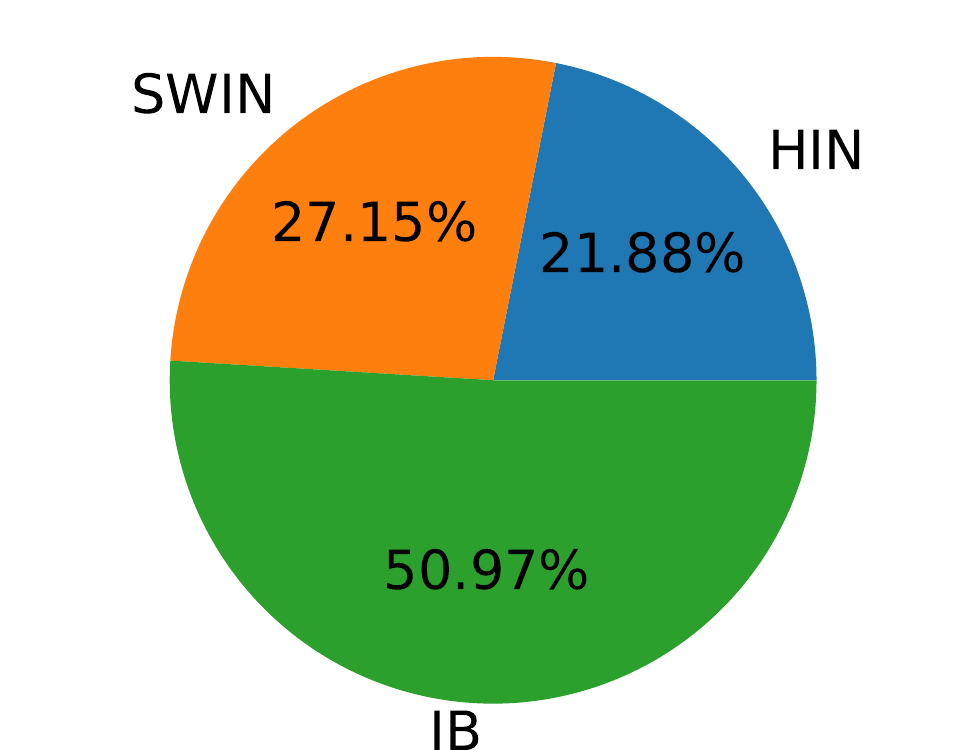}}%
	\subfloat[Part 2]{\includegraphics[width=0.3\linewidth]{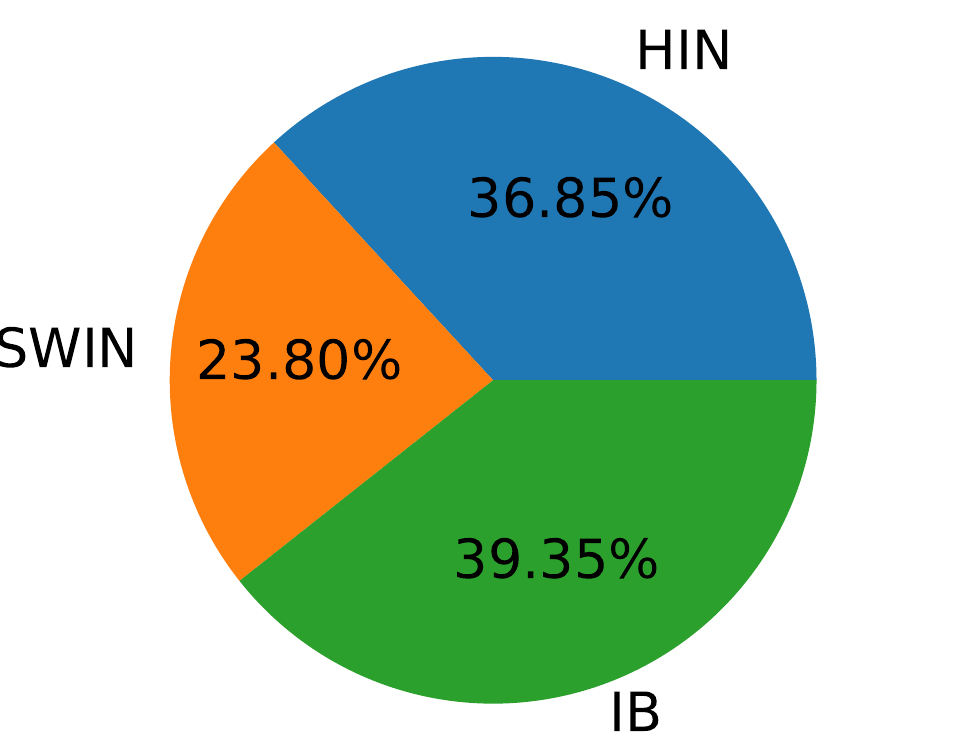}}%

	\caption{The percentages of specific denoising operators in `Part 0', `Part 1', and `Part 2'.}
	\label{fig:op_dis}
\end{figure}

\begin{figure}[tp]
	\centering
	\subfloat[HIN]{\includegraphics[width=0.3\linewidth]{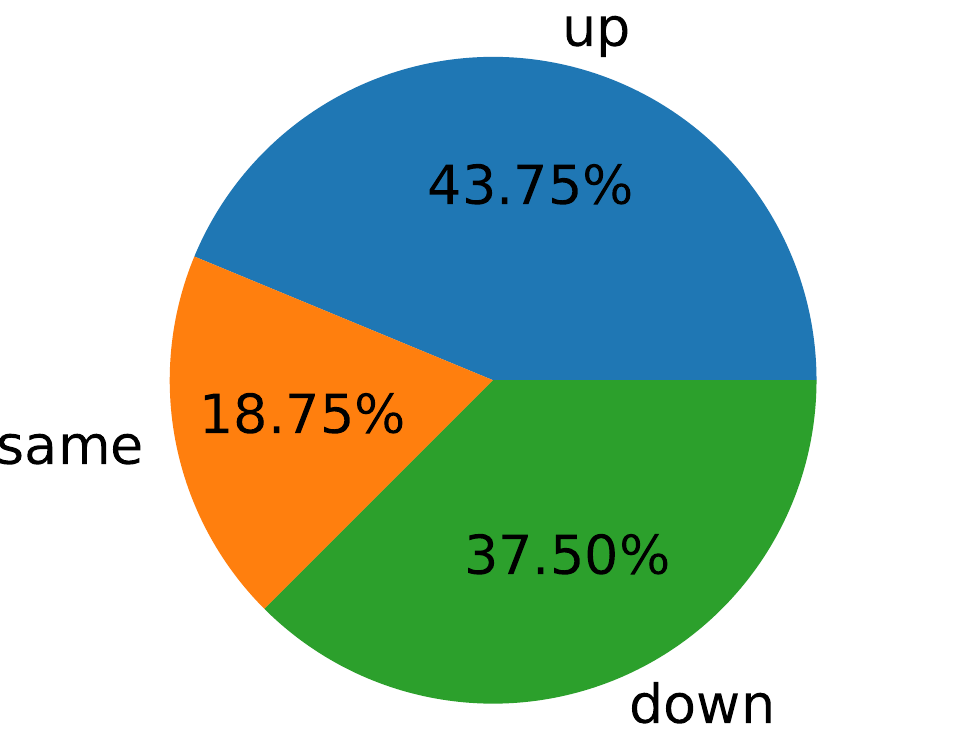}}%
	\subfloat[SWIN]{\includegraphics[width=0.3\linewidth]{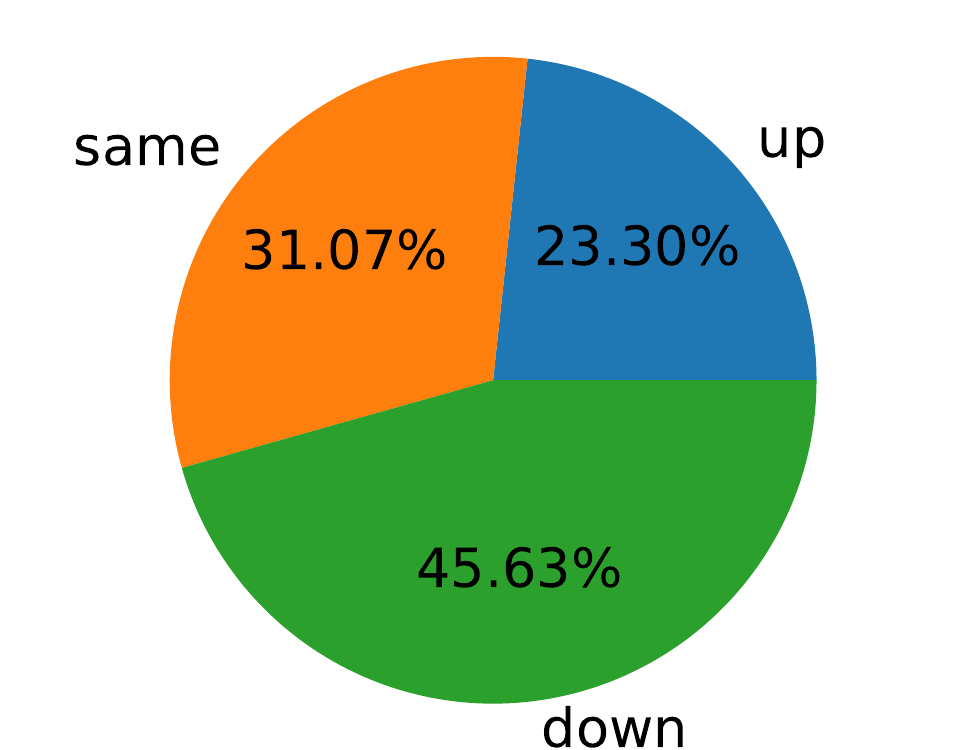}}%
	\subfloat[IB]{\includegraphics[width=0.3\linewidth]{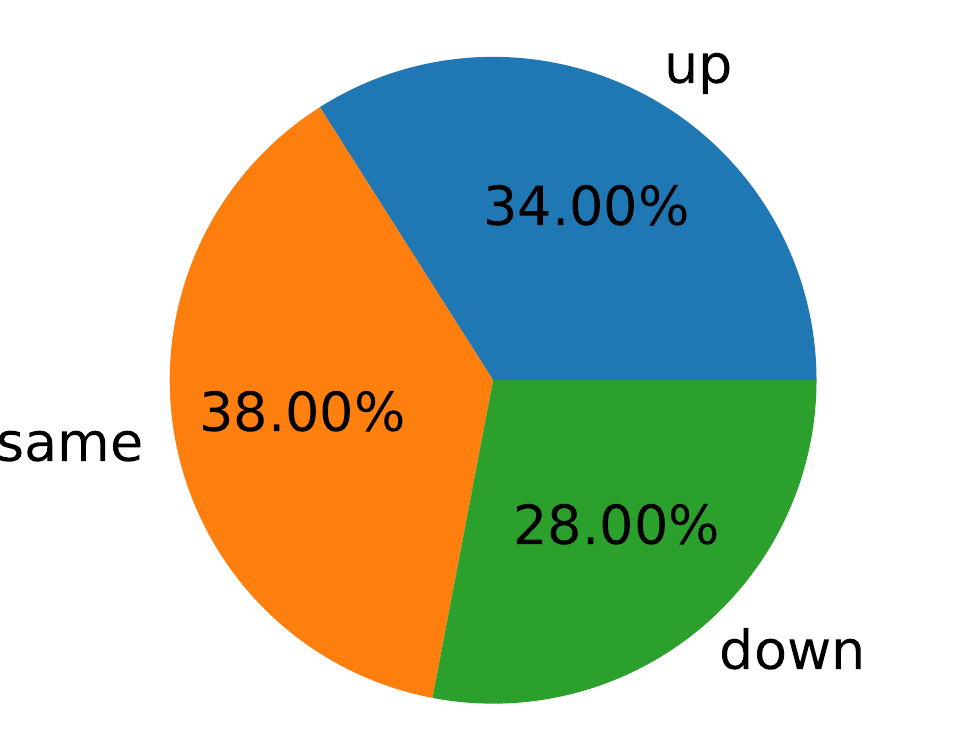}}%

	\caption{The percentages of resolutions which are required for the specific denoising operators. The `up' indicates ${\text{f}}_{up}({\boldsymbol x}_{\frac s2,\;l},\;{\boldsymbol x}_{s,\;m})$, the `same' indicates ${\boldsymbol x}_{s,\;l}$, and the `down' indicates ${\text{f}}_{down}({\boldsymbol x}_{2s,\;l})$. The legend can be found in Figure \ref{fig:framework}c.}
	\label{fig:op_resolu}
\end{figure}

\subsection{Discussion}
Compared with hand-crafted denoising methods, the NAS-based method could provide plenty of architectures for observation. We aim to summarize the preferences of operators and model complexity for further work. We select $200$ searched architectures with similar validation accuracy, and generate statistical results as below.

\textbf{Rates of selected operators.} As shown in Figure \ref{fig:hist}, the operator rates vary in different parts. 
The `skip' rate varies from $0.1$ in `Part 0' to $0.3$ in `Part 1' and `Part 2'. As we illustrate in the cell-level search in Section \ref{sec:Coarse-to-fine}, the `skip' modifies the depth of our searched architecture. The increment indicates that we may not need deeper architecture for the later parts. For convolution, the `Part 0' prefers convolutions with larger receptive field, i.e., `conv\_d3', while the others have no distinct preferences. 

\textbf{Percentages of denoising operators.} Since the convolutions and skip connection play a dominant role, the distinctions between the specific denoising operators are not obvious. We demonstrate the variation of the $3$ operators in Figure \ref{fig:op_dis}. The percentage of `SWIN' decreases from $55\% $ in `Part 0' to nearly $25\%$ in the others. The `SWIN' which could extract long-range information is essential in the former part. The `IB' tends to appear in the later parts.

\textbf{Percentages of resolutions.} To better utilize the specific denoising operators, we explore the compatible resolution of feature maps delivered to them in Figure \ref{fig:op_resolu}. Notably, only the operator that receives feature maps of $3$ resolutions is taken into account. 
The $3$ resolution feature maps are resized and weighted summed, then sent into the denoising operators. This process is shown in Figure \ref{fig:framework}c, and we analyze the rate of $\bm{\beta}$, i.e., $\{{\beta}_{\frac s2\rightarrow s}$, ${\beta}_{s\rightarrow s}, {\beta}_{2s\rightarrow s}\}$.
Surprisingly, the `SWIN' prefers feature maps which are down-sampled. 
Intuitively, the feature maps we send to the subsequent operator usually comes from the previous operator, which generates the feature maps at the same resolution. This reason might be that the down-sampled feature maps have more contextual information. The `SWIN' has a stronger capability to learn the context information and then gets better denoising performance. The `HIN' prefers up-sampled and down-sampled feature maps, while the `IB' has no preferences for resolution.

\textbf{Distributions of model complexity.} As our method provides architectures in different model complexities, we explore the allocation of model complexity for each part in Figure \ref{fig:big_small}. 
The width of curves shows that the complexity of `Models-Large' has a more concentrated distribution while the complexity of `Models-Small' has a long-ranged distribution, especially in `Part 1' and `Part 2'. Indicating that it is more difficult to design an excellent small model due to the inconsistency of complexity. 
From the quartile lines, we find the `Models-Large' tends allocate more complexity to the `Part 0' and `Part 2'. The `Model-Small' takes more attention to the `Part 1' and `Part 2'. This phenomenon suggests us focus on different parts for designing large or small models for image denoising.

The $3$ parts can be treated as the front, middle and rear of an architecture. For image denoising, we draw $3$ general guidelines for designing a model from the discussion:
\begin{enumerate}
	\item The front of the architecture should have a stronger learning capability to extract more discriminative features. This capability can be achieved by operators with a wider receptive field, like `SWIN' or `dilated convolution'.
	\item The `SWIN' and `HIN' prefer feature maps with more contextual information.
	\item In large models, more parameters or higher FLOPs should be taken for the front and rear of the architecture. The small models focus on the middle and rear of the architecture.
\end{enumerate}

\section{Conclusion}
We propose a coarse-to-fine search framework for image denoising. Our framework searches for an optimal architecture based on the network-level, the cell-level and the kernel-level space. For the search space, the coordinating search strategies provide the feasibility and extensibility for extending various denoising designs. In such a giant search space, the denoising prior-based regularization reduces the search difficulty while maintaining the diversity of our candidates. Together with inference time-based regularization, our framework balances the tradeoff between the model performance and model complexity. Based on our search framework, our searched architecture demonstrates the superiority on multiple synthetic and real-world datasets. Moreover, we discuss the phenomenon of searched architectures and hope to help design models for image denoising. 

\bibliographystyle{IEEEtran}
\bibliography{nas}
\end{document}